\documentclass[12pt]{article}

\usepackage{pgfplots}
\usetikzlibrary{decorations.pathmorphing}

\tikzset{snake it/.style={decorate, decoration=snake}}
\pgfplotsset{compat=1.10}
\usepgfplotslibrary{fillbetween}
\usetikzlibrary{patterns}
\usepackage{draft} 
\usepackage{hyperref}
\usepackage{graphicx,color,subfig}
\usepackage{cite}
\usepackage{mciteplus}
\usepackage{skak}
\allowdisplaybreaks
\usepackage{xcolor}
\usepackage{empheq}
\usepackage{tikz}
\DeclareFontFamily{OT1}{pzc}{}
\DeclareFontShape{OT1}{pzc}{m}{it}{<-> s * [1.10] pzcmi7t}{}
\DeclareMathAlphabet{\mathpzc}{OT1}{pzc}{m}{it}

\def\be#1\ee{\begin{align}#1\end{align}}



\newcommand{\jinwei}[1]{{\bf {\color{red} JC:}} {{\color{blue}\it#1}}}
\newcommand{\david}[1]{{\bf {\color{red} DK:}} {{\color{blue}\it#1}}}
\newcommand{\IE}[0]{{\textit{i.e. }}}
\newcommand{\EG}[0]{{\textit{e.g. }}}

\begin{document}

\unitlength = .8mm

\begin{titlepage}

\begin{center}

\hfill \\
\hfill \\
\vskip 1cm

\title{On Time -- Dependent Backgrounds \\In $1+1$ Dimensional String Theory\Huge}
\author{Bruno Balthazar$^{1,2}$, Jinwei Chu$^1$,
David Kutasov$^1$}
\address{$^1$Kadanoff Center for Theoretical Physics and Enrico Fermi Institute\\ University of Chicago, Chicago IL 60637\\~\\ $^2$Center for Cosmology and Particle Physics\\ New York University, New York NY 10003}
\vskip 1cm

\email{brunobalthazar@nyu.edu, jinweichu@uchicago.edu, dkutasov@uchicago.edu}

\end{center}

\abstract{In perturbative string theory, one is generally interested in asymptotic observables, such as the S-matrix in flat spacetime, and boundary correlation functions in anti-de Sitter spacetime. However, there are backgrounds in which such observables do not exist. We study examples of such backgrounds in $1+1$ dimensional string theory. In these examples, the Liouville wall accelerates and can become spacelike in the past and/or future. When that happens, the corresponding null infinity, at which the standard scattering states are defined, is shielded by the Liouville wall. We compute scattering and particle production amplitudes in these backgrounds in the region in parameter space where the wall remains timelike, and discuss the continuation of this picture to the spacelike regime. We also discuss the physics from the point of view of the dynamics of free fermions in backgrounds with a time-dependent Fermi surface.
}

\vfill

\end{titlepage}

\eject

\begingroup
\hypersetup{linkcolor=black}
\tableofcontents
\endgroup
\section{Introduction}\label{intro}

Since its inception in the late 1980's, string theory in a $1+1$ dimensional spacetime has played an important role in the development of the subject (see \EG \cite{Seiberg:1990eb,Klebanov:1991qa,Ginsparg:1993is,Jevicki:1993qn,Nakayama:2004vk,Martinec:2004td,Anninos:2020ccj} for reviews). In particular, its dual description in terms of matrix quantum mechanics in a double scaling limit was an early example of holography, a precursor to the AdS/CFT correspondence \cite{Maldacena:1997re,Gubser:1998bc,Witten:1998qj,Aharony:1999ti} and Little String Theory \cite{Aharony:1998ub,Aharony:1999ks,Kutasov:2001uf}. The fact that the theory is solvable allows one to use it to test general ideas in string theory and holography in a setting where explicit calculations can be performed. An example is the study of non-perturbative effects (in the string coupling); see \EG  \cite{Balthazar:2019rnh,Balthazar:2019ypi,Sen:2019qqg,Sen:2020oqr,Sen:2020ruy,Sen:2020eck,Sen:2021qdk,DeWolfe:2003qf,Balthazar:2022apu,Chakravarty:2022cgj,Sen:2022clw,Eniceicu:2022xvk,Alexandrov:2023fvb} for some recent discussions.  

The theory was originally formulated in a static background, where a natural set of observables is given by S-matrix elements for scattering of (massless) ``tachyons''. It is known that the theory is solvable in time-dependent backgrounds as well (see \EG \cite{Moore:1992gb,Moore:1992ga,Dijkgraaf:1992hk,Hsu:1992cm,Kazakov:2000pm,Alexandrov:2002fh,Alexandrov:2002pz,Alexandrov:2003uh,Karczmarek:2003pv,Das:2004hw,Das:2004aq,Das:2007vfb}), however in some of these backgrounds the nature of the observables is less clear. The main goal of this paper is to discuss the physics of a class of such backgrounds, and try to learn from them how to treat backgrounds with similar features in more realistic settings in higher dimensions, such as backgrounds with cosmological singularities.

\subsection{The standard background}\label{tind}

To set the stage, we start with a brief review of the situation in the standard background \cite{Seiberg:1990eb,Klebanov:1991qa,Ginsparg:1993is,Jevicki:1993qn,Nakayama:2004vk,Martinec:2004td,Anninos:2020ccj}. In Euclidean two dimensional spacetime with (Euclidean) worldsheet metric $\hat g$, it is described by the action
\begin{equation}
\label{Sws}
S_E=\frac{1}{4\pi}\int d^2z\sqrt{\hat{g}}\left[\left(\hat\nabla X\right)^2+\left(\hat{\nabla}\phi\right)^2+2\phi R(\hat{g})-4\pi\mu \phi e^{2 \phi}\right].
\end{equation}
The linear dependence of the dilaton on $\phi$ leads to a string coupling that behaves like $g_s(\phi)\sim \exp(2\phi)$. One can think of $\phi$ as the conformal factor of a dynamical metric, and of \eqref{Sws} as a conformal gauge description of worldsheet gravity coupled to a massless scalar field, $X$.

The worldsheet theory \eqref{Sws} becomes free in the region $\phi\to-\infty$, where the string coupling $g_s(\phi)$ goes to zero. This region is the boundary of the two dimensional spacetime labeled by $(X,\phi)$. Conversely, as $\phi$ increases, the string coupling increases, and one might think that perturbative string theory breaks down due to strong coupling effects associated with the large positive $\phi$ region. However, the cosmological constant term in the action \eqref{Sws} gives rise to a potential for $\phi$ that prevents it from exploring this region.\footnote{Here $\mu$ is taken to be positive, and as discussed below, the form of the potential in \eqref{Sws} is only valid for large negative $\phi$. As $\phi$ increases, the potential receives corrections, so that it goes to infinity as $\phi\to\infty$.} This potential is often referred to as the {\it Liouville wall}. It leads to the fact that in the background \eqref{Sws} the $g_s$ expansion is essentially a $1/\mu$ expansion.  

Physical observables in the background \eqref{Sws} are correlation functions of vertex operators characterized by their behavior near the boundary. A large set of such operators corresponds to modes of the massless ``tachyon'' field, described by the vertex operators\footnote{Here and below we take the worldsheet metric $\hat g$ to be flat, \IE $z$ is a coordinate on the complex plane with flat metric.}
\begin{equation}
\label{vo}
    T_p\simeq \frac{\Gamma(|p|)}{\Gamma(1-|p|)}\int d^2ze^{(2-|p|)\phi}e^{ipX}~.
\end{equation}
In \eqref{vo} we chose to normalize the operators in a conventional way; see \EG \cite{Kutasov:1991pv,Ginsparg:1993is}. The $\simeq$ means that \eqref{vo} describes the behavior of the operators as $\phi\to-\infty$; the finite $\phi$ form of the operators \eqref{vo} is more complicated. The physical observables are correlation functions of the operators $T_p$,
\ie
\label{corf}
\langle T_{p_1}T_{p_2}\cdots T_{p_l}\rangle\ .
\fe
These correlation functions are uniquely determined by the asymptotic form \eqref{vo}.

An important feature of the operators \eqref{vo} is that they are non-normalizable. Indeed, their wavefunctions behave in the limit $\phi\to-\infty$ like $\Psi_p(X,\phi)\sim e^{(ipX-|p|\phi)}$. Thus, adding them to the worldsheet Lagrangian corresponds to a deformation of the Lagrangian of the spacetime theory. We note, for future reference, that the wavefunctions $\Psi_p$ have the property that for positive (negative) $p$ they depend on the (anti-)holomorphic variable $\phi\mp iX$.

The worldsheet field $X$ is often taken to be compact, $X\sim X+2\pi R$, \EG to study the theory at finite temperature. In that case the momentum $p$ is quantized, $pR\in Z$, and there are also winding modes. The latter will not play a role in our discussion below. They are related to momentum modes by T-duality, which can be used to relate their correlation functions to those of momentum modes, \eqref{corf}. 

The cosmological term in the worldsheet action \eqref{Sws} can be written as $\mu T_0$. The prefactor in \eqref{vo} has a pole at $p=0$; this pole is responsible for the factor of $\phi$ in front of the exponential in \eqref{Sws}.   Differentiating the path integral w.r.t. $\mu$ gives an insertion of $-T_0$. We will use this fact in the calculation of the correlation functions \eqref{corf}.

A property of these correlation functions that will be useful below is their dependence on $\mu$, known as KPZ scaling. By analyzing the behavior of \eqref{corf} under a shift of $\phi$, one can show that (to leading order in $g_s$)
\ie
\label{KPZs}
\langle T_{p_1}T_{p_2}\cdots T_{p_l}\rangle\sim \mu^a ,
\fe
with $a$ determined by 
\ie
\label{forms}
\sum_{j=1}^l \left(2-|p_j|\right)+2a=4~.
\fe
The coefficient of $\mu^a$ in \eqref{KPZs} contains the non-trivial dependence of \eqref{corf} on the momenta $p_j$. In order to determine it, one needs to solve the worldsheet theory \eqref{Sws}.

\begin{figure}[h]
\centering
 \includegraphics[width=0.45\textwidth]{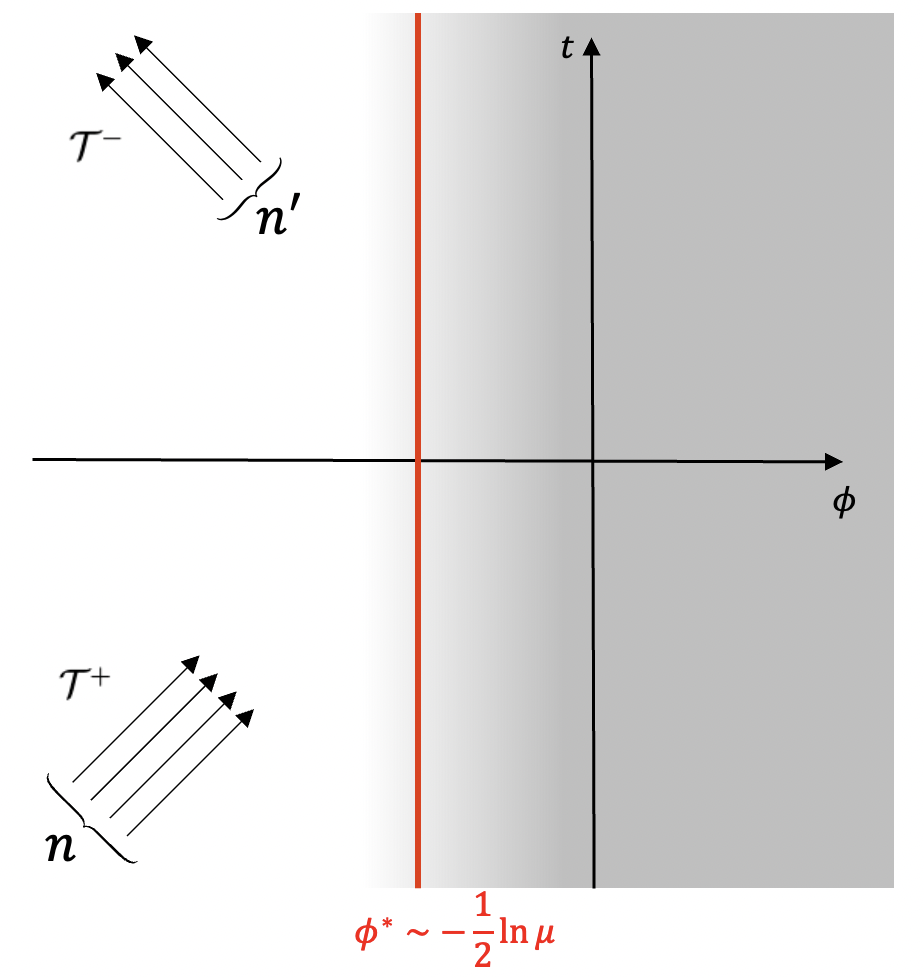}
 \caption{The correlation functions \eqref{corS} describe scattering processes, where $n$ right-moving tachyons $\mathcal{T}^+$ come in from the boundary $\phi\to-\infty$ and scatter off the Liouville wall (depicted in red) into $n'$ left-moving tachyons $\mathcal{T}^-$ that go back towards the boundary.} 
\label{1powa}
\end{figure}

The Lorentzian analog of \eqref{Sws} is obtained by taking $X\to -it$, which gives
\begin{equation}
\label{SwsL}
S_L=\frac{1}{4\pi}\int d^2z\left(-\left(\nabla t\right)^2+\left(\nabla\phi\right)^2-4\pi\mu \phi e^{ 2 \phi}\right)\ .
\end{equation}
The target spacetime is now $1+1$ dimensional, with a spacelike linear dilaton and a Liouville wall in the spatial $(\phi)$ direction, which as before shields the strong coupling region. This is depicted in figure \ref{1powa}, where the red line is the Liouville wall, defined as the surface along which the Liouville potential $V(\phi)$ is of order one.\footnote{This wall is soft, in the sense that particles with higher energy penetrate it to larger $\phi$; see \EG \cite{Seiberg:1992bj} for a discussion.} The shaded region is shielded by the Liouville potential.

The natural Lorentzian observables are obtained by taking $|p|\to -i\omega$ ($\omega>0$) in \eqref{vo}, \footnote{Each of the vertex operators $\mathcal{T}^\pm_\omega$ is in fact a linear combination of incident and reflected waves, as is standard in quantum mechanics. The relative coefficient between the incident and reflected waves is the reflection coefficient.}
\begin{equation}
\label{voL}
    \mathcal{T}^\pm_\omega\simeq \frac{\Gamma(-i\omega)}{\Gamma(1+i\omega)}\int d^2ze^{(2+i\omega)\phi}e^{\mp i\omega t}~.
\end{equation}
The vertex operators \eqref{voL} describe particles with energy $\omega$ propagating in the background \eqref{SwsL}. $\mathcal{T}^+_\omega$ corresponds to an incoming particle, moving to the right, towards the wall, while $\mathcal{T}^-_\omega$ corresponds to an outgoing, left-moving, particle with energy $\omega$, moving away from the wall (see figure \ref{1powa}). The corresponding wavefunctions $\Psi^\pm_\omega(t,\phi)\sim e^{i\omega(\phi\mp t)}$ are $\delta$-function normalizable, as appropriate for describing asymptotic particle states.

Correlation functions of the form 
\begin{equation}
\label{corS}
\langle\prod_{j=1}^{n}\mathcal{T}^+_{\omega_j}\prod_{l=1}^{n'}\mathcal{T}^-_{\omega'_l}\rangle 
\end{equation}
describe scattering amplitudes of $n$ tachyons with energies $\{\omega_j\}$ to $n'$ tachyons with energies $\{\omega'_l\}$  (see figure \ref{1powa}). Time translation invariance of the background implies that energy is conserved in such processes, $\sum_{j=1}^n\omega_j=\sum_{l=1}^{n'}\omega'_l$.

As before, the $\mu$ dependence of \eqref{corS} is determined by KPZ scaling. The analog of  \eqref{KPZs}, \eqref{forms} is
\begin{equation}
\label{corSL}
\langle\prod_{j=1}^n\mathcal{T}^+_{\omega_j}\prod_{l=1}^{n'}\mathcal{T}^-_{\omega'_l}\rangle \sim \mu^a~,
\end{equation}
with 
\ie
\label{formrlor}
\sum_{j=1}^n \left(2+i\omega_j\right)+\sum_{l=1}^{n'} \left(2+i\omega_l'\right)+2a=4~.
\fe
In particular, the dependence of the amplitudes \eqref{corS} on $\omega_j$ is via the phase  $\mu^{-\frac i2\omega_j}=e^{-\frac i2\omega_j\ln\mu}$, and similarly the dependence on $\omega_l'$ is via the phase $e^{-\frac i2\omega_l'\ln\mu}$. Looking back at the form of the vertex operators $\mathcal{T}^\pm_\omega$, \eqref{voL}, we see that this factor has a simple interpretation -- it indicates that the scattering process described by the correlation function \eqref{corS} happens in the vicinity of the wall, $\phi^*\sim -\frac12\ln\mu$ (see figure \ref{1powa}), and thus is sensitive to the value of the vertex operators at that location. 

As in the Euclidean case, the dynamical information about the scattering processes \eqref{corS} is contained in the coefficient of $\mu^a$ in \eqref{corS}. In principle, one can obtain it by calculating the scattering amplitudes \eqref{corS} using the worldsheet description \eqref{SwsL}, but in practice this is difficult for general $n$ and $n'$, due to the interacting nature of the worldsheet theory. The dual matrix model allows one to calculate these amplitudes much more efficiently. It also allows one to calculate quantum corrections to the leading terms, which from the worldsheet perspective come from higher genus contributions to the correlation functions \eqref{corS}. We will use some of the matrix model results below.

\subsection{Time-dependent backgrounds}\label{introtd}

As mentioned above, our main interest will be in time-dependent backgrounds, that are obtained by deforming \eqref{SwsL}. Such backgrounds were discussed in the past, \EG in \cite{Moore:1992gb,Moore:1992ga,Dijkgraaf:1992hk,Hsu:1992cm,Kazakov:2000pm,Alexandrov:2002fh,Alexandrov:2002pz,Alexandrov:2003uh,Karczmarek:2003pv,Das:2004hw,Das:2004aq,Das:2007vfb}, and we will use some of their results. A special case of these backgrounds was recently revisited in \cite{Rodriguez:2023kkl,Rodriguez:2023wun,Collier:2023cyw}.

The particular deformations we will consider correspond to adding to the worldsheet action the terms\footnote{Here and below, we take $p>0$, without loss of generality.}  
\begin{equation}
\label{defL}
\delta S_L=\lambda_+T_p+\lambda_-T_{-p}=\frac{\Gamma(p)}{\Gamma(1-p)}\int d^2ze^{(2-p)\phi}\left(\lambda_+e^{pt}+\lambda_-e^{-pt}\right).
\end{equation}
The couplings $\lambda_\pm$ are taken to be real, so that the worldsheet action remains real after deformation. If $\lambda_\pm$ are both non-zero, one can take them to satisfy $\lambda_+=\pm\lambda_-$ by shifting the origin of time. We will mostly focus on the case where one of the two couplings vanishes, in which case the remaining coupling controls the time at which the perturbation \eqref{defL} becomes important. Obviously, for $\lambda_-=0$ the perturbation \eqref{defL} becomes important in the future, while for $\lambda_+=0$ it becomes important in the past.

We see from \eqref{defL} that there is a qualitative difference between the cases $0<p<2$ and $p>2$. In the former case, the perturbation goes to zero near the boundary of spacetime $\phi\to-\infty$ (for fixed $t$). In the latter, it grows as we approach the boundary. The origin of this behavior is well understood. For $p<2$, the operators $\exp(\pm pt)$ are relevant, and thus, after coupling to $\phi$, their effective coupling in \eqref{defL} goes to zero in the UV region $\phi\to-\infty$. Conversely, for $p>2$ these operators are irrelevant, and modify the UV behavior. We will restrict to the case $p<2$ below.

\begin{figure}[h!]
\centering
 {\subfloat[\label{2powa1}]{\includegraphics[width=0.4\textwidth]{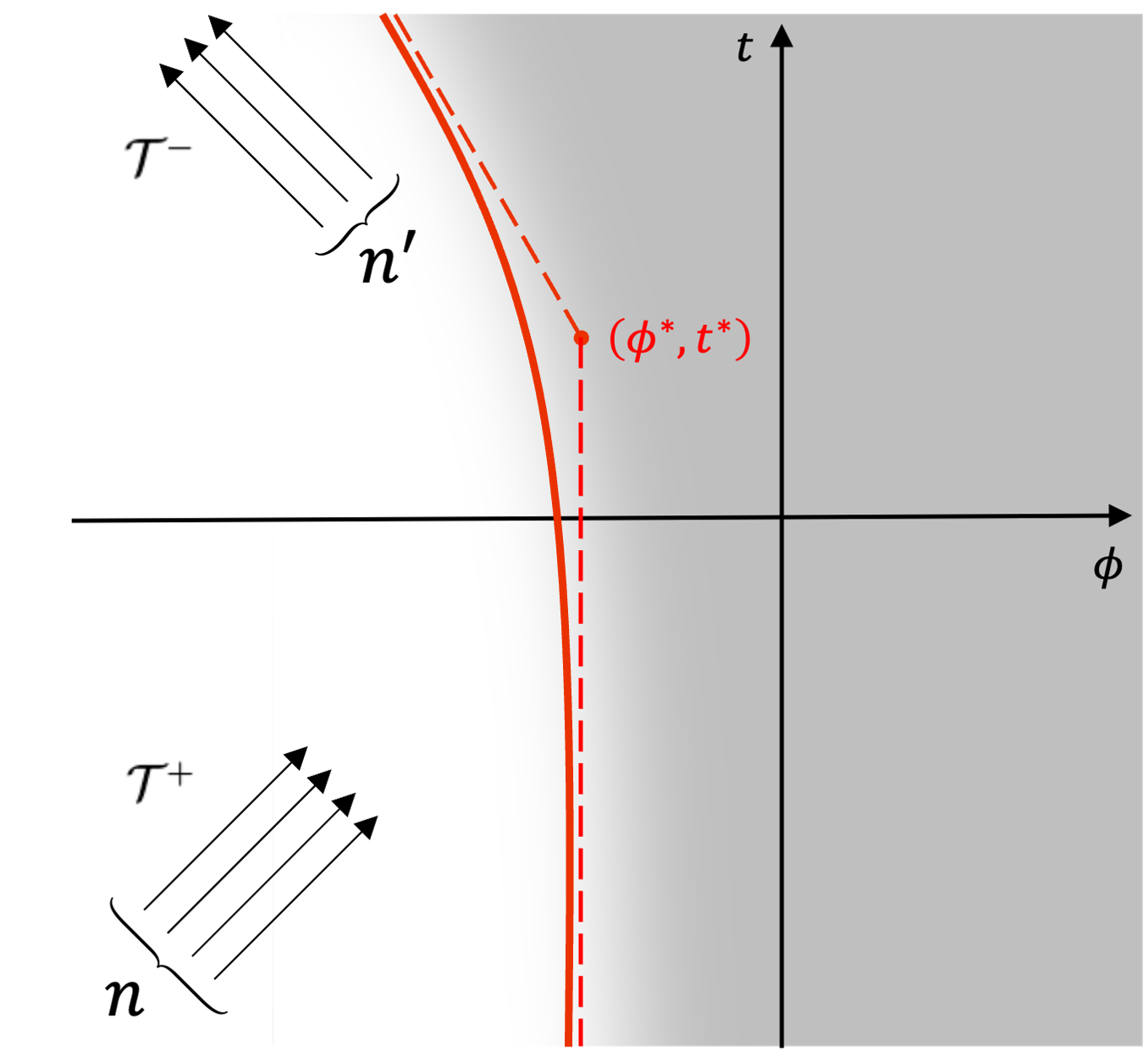}}}~
 \hspace{1 cm}
 {\subfloat[\label{2powa2}]{\includegraphics[width=0.4\textwidth]{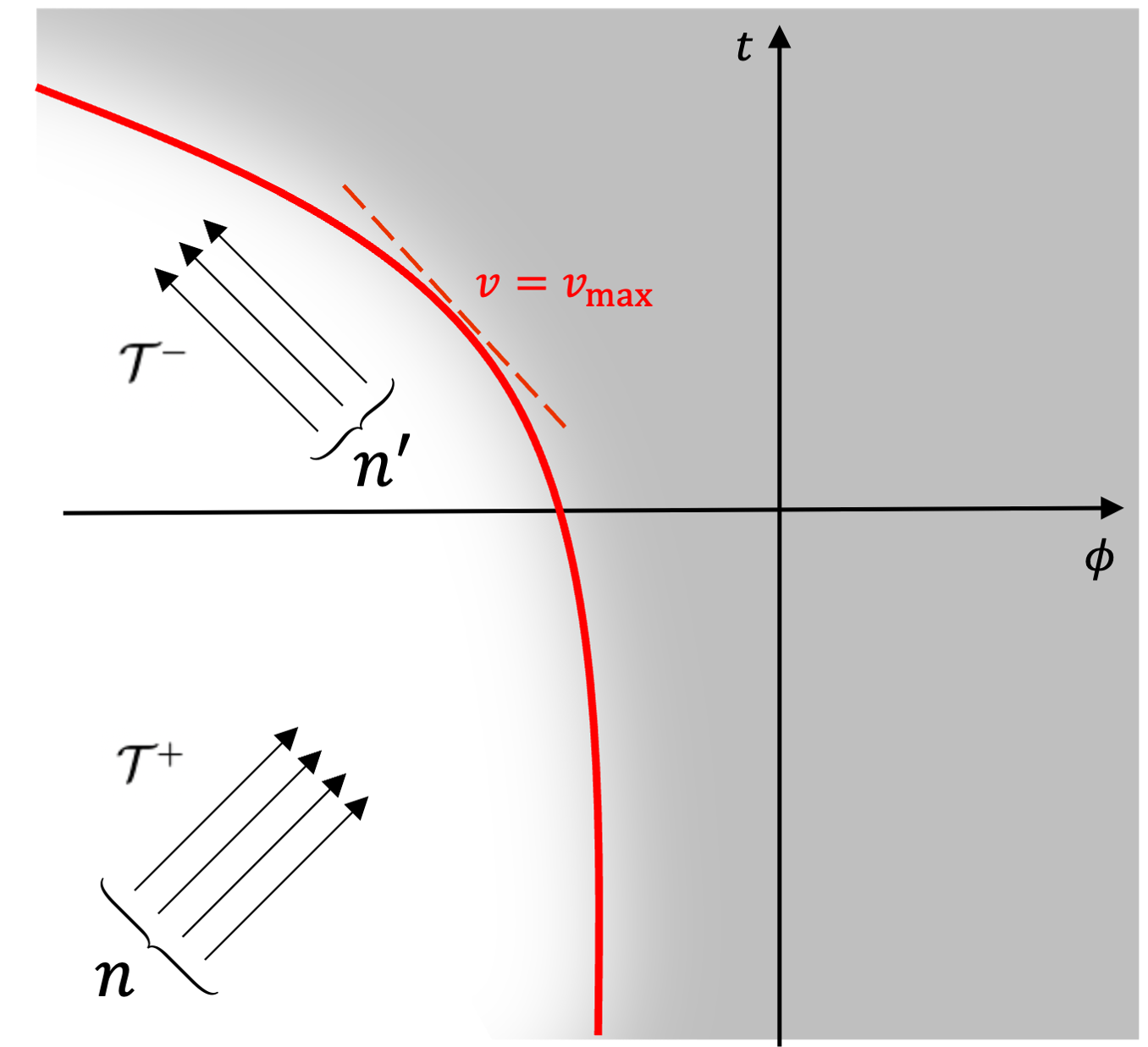}}} 
  \hspace{0.5 cm}
 \caption{In the presence of the perturbation \eqref{wspot}, the Liouville wall becomes time-dependent. As $t\to-\infty$, it approaches the static wall of the unperturbed system, while as $t\to +\infty$, its velocity  approaches $-\frac{p}{2-p}$. For $p<1$, the Liouville wall remains timelike for all $t$, (a), while for $p>1$ it eventually becomes spacelike, (b).}
 \label{2powa}
\end{figure}

The addition of the perturbation \eqref{defL} to the worldsheet Lagrangian \eqref{SwsL} modifies the worldsheet potential. From now on, we will focus on the case $\lambda_-=0$, unless explicitly stated otherwise. The worldsheet potential takes in this case the form (at large negative $\phi$)
\begin{equation}
\label{wspot}
V_{\rm ws}(t,\phi)=-\mu \phi e^{2 \phi}+\hat\lambda_+e^{(2-p)\phi+pt}~,
\end{equation}
where $\hat\lambda_+=\lambda_+\frac{\Gamma(p)}{\Gamma(1-p)}$. 
The situation is described in figure \ref{2powa}, which generalizes figure \ref{1powa} to $\hat\lambda_+>0$. The solid red line in figure \ref{2powa} is the surface $V(t,\phi)\sim 1$, which can be thought of as a {\it time-dependent Liouville wall}. 

For early time $(t\to-\infty)$, the perturbation proportional to $\hat\lambda_+$ in \eqref{wspot} goes to zero, and we recover the static Liouville wall described in the previous subsection. For late time $(t\to\infty)$, the velocity of the modified Liouville wall approaches a finite constant, as indicated in figure \ref{2powa}. The transition between the two regimes occurs around the point $(\phi,t)=(\phi^*,t^*)$,  
\ie
\label{ptstar}
(\phi^*,t^*)=\Big(-\frac{1}{2}\ln \mu,-\frac{1}{p}\ln \hat{\lambda}_++\frac{2-p}{2p}\ln \mu\Big)
\fe
depicted in figure \ref{2powa}. This is the regime in which the Liouville wall accelerates from its initial to its final velocity.

As is evident from the figure, there is a qualitative difference between the cases $0<p<1$ and $1<p<2$. In the former case, exhibited in figure \ref{2powa1}, the trajectory of the Liouville wall is timelike for all $t$. On the other hand, for $1<p<2$, figure \ref{2powa2}, the Liouville wall is spacelike at large $t$, \IE it moves faster than light at late times. Note that this is not inconsistent with special relativity, since the Liouville wall is not a dynamical object. It is rather a non-normalizable background, and thus cannot be used to propagate information faster than light. The dynamical field -- the tachyon -- is massless, and can be used to transmit information at the speed of light.

There is an important subtlety in the above discussion that will play a role in our analysis. In the holographic map between $1+1$ dimensional string theory and matrix quantum mechanics, the tachyon field on the worldsheet, whose momentum modes are given by the exponentials in equations \eqref{vo}, \eqref{voL}, and its matrix model analog, are related by a momentum dependent factor -- the ratio of Gamma functions in these formulae. This ratio can be thought of as due to the non-zero modes of the worldsheet fields $(\phi,t)$. Therefore, the deformation \eqref{defL}, that gives rise to the worldsheet interaction $\int d^2z V_{\rm ws}(t(z,\bar z),\phi(z,\bar z))$, \eqref{wspot}, yields the  potential 
\begin{equation}
\label{sppot}
V_{\rm st}(t,\phi)=\mu T_0+\lambda_+T_p=\mu e^{2 \phi}+\lambda_+e^{(2-p)\phi+pt}~
\end{equation}
for the zero modes of the worldsheet fields $(\phi(z,\bar z), t(z,\bar z))$. The time-dependent Liouville wall is described by the equation $V_{\rm st}(t,\phi)=1$. 

The qualitative discussion above of the worldsheet potential $V_{\rm ws}$ mostly goes through when it is replaced by $V_{\rm st}$, but there are a couple of important changes. First, the time-dependent term in \eqref{sppot} is positive for $\lambda_+>0$, whereas in \eqref{wspot} it is $\hat\lambda_+$ that must be positive. For $0<p<1$ the two notions coincide, but for $1<p<2$ they differ. We will take the point of view that the spacetime potential $V_{\rm st}$ is the important one for the dynamics of the tachyon field, and thus take $\lambda_+$ to be positive. We will see later that this gives a coherent picture of the physics. The second change is that the point $(\phi^*, t^*)$ in figure \ref{2powa} is still given by \eqref{ptstar}, but with $\hat\lambda_+$ replaced by $\lambda_+$,
\ie
\label{ptstarst}
(\phi^*,t^*)=\Big(-\frac{1}{2}\ln \mu,-\frac{1}{p}\ln \lambda_++\frac{2-p}{2p}\ln \mu\Big).
\fe
This too will play a role in our analysis.

The form of the time-dependent Liouville wall has an important impact on the observables in this background. Past lightlike infinity is part of the boundary for all $p<2$, and therefore we can define the observables $\mathcal{T}^+_\omega$ in figure \ref{2powa} for non-zero $\lambda_+$, $p$. However, future lightlike infinity is only part of the boundary for $p<1$. Hence, the outgoing particles $\mathcal{T}^-_\omega$ can only be defined as asymptotic observables in that regime. Thus, for $p<1$, we can compute scattering amplitudes of the sort \eqref{corS} with any $n,n'\ge 0$, as depicted in figure \ref{2powa}. However, for $1<p<2$ we cannot define amplitudes with $n'>0$, since the region where the operators $\mathcal{T}^-_\omega$ are to be defined is shielded by the potential \eqref{sppot}. 

This issue is especially significant when both $\lambda_+$ and $\lambda_-$ are positive and $1<p<2$. In that case it seems like there are no good observables, since 
both $\mathcal{T}^+_\omega$ and $\mathcal{T}^-_\omega$ appear not to exist, for the same reason as above. We will mainly focus on the case $\lambda_-=0$ and $p<1$ below, but will comment on $p>1$ and $\lambda_->0$ at various points in our analysis. 

The fact that the background \eqref{defL} with $\lambda_-=0$ is time-dependent implies that energy is not conserved. For example, we can consider amplitudes of the form 
\begin{equation}
\label{absor}
\big\langle\prod_{j=1}^n\mathcal{T}^+_{\omega_j}\big\rangle~,
\end{equation}
where $n$ particles are incident on the wall and are absorbed by the time-dependent background. For $p<1$ we can also consider the amplitudes 
\begin{equation}
\label{emit}
\big\langle\prod_{l=1}^{n'}\mathcal{T}^-_{\omega'_l}\big\rangle~,
\end{equation}
corresponding to the creation of $n'$ outgoing particles in the time-dependent Liouville potential \eqref{wspot}. And, of course, we can consider general amplitudes of the form \eqref{corS} with any non-negative $(n,n')$, and any energies $(\omega_j,\omega'_l)$.

In the case where the Liouville wall is static, KPZ scaling \eqref{corSL}, \eqref{formrlor} provides insight into the dynamics -- it suggests that the scattering processes of figure \ref{1powa} generically happen in the vicinity of the Liouville wall. It is thus interesting to generalize the scaling analysis to the case of figure \ref{2powa}. For example, for the amplitude \eqref{absor} one has
\begin{equation}
\label{KPZabsor}
\big\langle\prod_{j=1}^n\mathcal{T}^+_{\omega_j}\big\rangle\sim \mu^a\lambda_+^b\;.
\end{equation}
The powers $a,b$ can be determined by analyzing the behavior of the amplitude \eqref{KPZabsor} under shifts of $\phi$ and $t$. This gives
\ie
\label{phimom}
\sum_{j=1}^n\left(2+i\omega_j\right)+2a+\left(2-p\right)b=4\ ,
\fe
and
\ie
\label{Xmom}
-i\sum_{j=1}^n\omega_j+bp=0\ ,
\fe
respectively. The solution of these equations is
\ie
\label{rsplus}
a=-\frac{i}{p}\sum_{j=1}^n\omega_j-n+2~,\quad b=\frac{i}{p}\sum_{j=1}^n\omega_j\ .
\fe
Plugging \eqref{rsplus} into \eqref{KPZabsor}, we see that the dependence of the correlator on each of the energies $\omega=\omega_j$ is via the phase factor 
\ie
\label{factorplus}
\left(\frac{\lambda_+}{\mu}\right)^{\frac{i\omega}{p}}=\exp\left(\frac{i\omega}{p}\ln\frac{\lambda_+}{\mu}\right)
\fe
These factors have a similar interpretation to the one discussed for the static background after eq. \eqref{formrlor}. Evaluating the vertex operator $\mathcal{T}^+_{\omega}$ \eqref{voL} at the location $(\phi^*, t^*)$ near which the Liouville wall accelerates (see figure \ref{2powa}), given in \eqref{ptstarst}, we find that at that location the vertex operator is given by $1/\mu$ multiplying the phase factor \eqref{factorplus}. The former is the factor of $g_s$ that relates a vertex operator to the wavefunction, while the latter implies that the amplitude \eqref{KPZabsor} is dominated by the region near $(\phi^*, t^*)$.  

In other words, as one would have expected, the time-dependent Liouville wall of figure \ref{2powa} can best absorb incident tachyons in the region where its velocity changes from zero to a finite value. The precise location and width of this region depends on the parameters, $\lambda_+$, $\mu$, and the energies of the particles $\omega_j$. And, as mentioned above, to actually compute this amplitude one needs to evaluate the coefficient of $\mu^a\lambda_+^b$ in \eqref{KPZabsor}. We will address this problem in the next sections. 

The above discussion can be repeated for the amplitude \eqref{emit}, which corresponds to a process where no particles are incident on the time-dependent Liouville wall, and $n'$ particles with energies $\{\omega'_l\}$ are emitted by the wall. As before, we have 
\begin{equation}
\label{KPZemit}
\big\langle\prod_{l=1}^{n'}\mathcal{T}^-_{\omega_l'}\big\rangle\sim \mu^a\lambda_+^b \ ,
\end{equation}
with $(a,b)$ determined by the conditions
\ie
\sum_{l=1}^{n'}\left(2+i\omega_l'\right)+2a+\left(2-p\right)b=4\ ,
\fe
and
\ie
i\sum_{l=1}^{n'}\omega_l'+bp=0\ .
\fe
The solution is
\ie
a=i\frac{1-p}{p}\sum_{l=1}^{n'}\omega_l'-n+2~,\quad b=-\frac{i}{p}\sum_{l=1}^{n'}\omega_l'\ .
\fe
The analog of the phase factor \eqref{factorplus} is in this case 
\ie
\label{factorminus}
\left(\lambda_+\mu^{p-1}\right)^{-\frac{i\omega}{p}}=\exp\left[-\frac{i\omega}{p}\ln(\lambda_+\mu^{p-1})\right].
\fe
The phase factor \eqref{factorminus} has the same interpretation as before -- it is obtained by evaluating the vertex operator $\mathcal{T}^-_{\omega}$ \eqref{voL} at the location $(\phi^*, t^*)$. As expected, we conclude that the emission of tachyons from the time-dependent Liouville wall of figure \ref{2powa} is centered in the region where the wall accelerates.   

We finish this subsection with some comments:
\begin{itemize}
\item Looking back at figure  \ref{2powa}, one can ask whether there are processes in which incoming tachyons are reflected from the Liouville wall far from the region where the wall accelerates, \EG at $t\ll t^*$. Such processes are of course possible, but as is clear from the figure, they are insensitive to $\lambda_+$, and in particular conserve energy. What we found above is that amplitudes that {\it violate} energy conservation are in general due to physics in the vicinity of $(\phi^*, t^*)$. 
\item In the discussion above we took either $n$ or $n'$ to vanish, but it is easy to generalize to any non-negative $n$ and $n'$. As expected, one finds that energy violating processes of the sort depicted in figure \ref{2powa} are dominated by the region where the Liouville wall accelerates.
\item Since the backgrounds we consider are not time translation invariant, the amplitudes we compute below receive, in general, contributions from disconnected diagrams. We will focus on connected diagrams, but it is easy to include disconnected ones.
\item As mentioned above, to compute amplitudes like \eqref{absor}, \eqref{emit}, we need to calculate the coefficients of $\mu^a\lambda_+^b$ in \eqref{KPZabsor}, \eqref{KPZemit}. In general, this requires the use of the matrix model. In the next subsection we explain our strategy for doing this calculation.  
    \end{itemize}

\subsection{Strategy of the calculation}

We will take an approach to this problem that proved fruitful in studying time-dependent solutions in open string theory (see \EG \cite{Sen:2004nf} for a review). We start with the Euclidean problem \eqref{Sws}, and add to the worldsheet action the term
\begin{equation}
\label{defE}
\delta S_E=\lambda_+T_p+\lambda_-T_{-p}=\frac{\Gamma(p)}{\Gamma(1-p)}\int d^2ze^{(2-p)\phi}\left(\lambda_+e^{ipX}+\lambda_-e^{-ipX}\right).
\end{equation}
As mentioned above, we will mostly consider the case $\lambda_-=0$. At first sight this looks problematic, since the worldsheet action is not real unless $\lambda_-=\lambda_+^*$, however this is standard in field theory (and string theory). We can study the theory with complex action \eqref{defE}, with the understanding that we are really interested in the Lorentzian theory obtained by taking $X\to -it$. In that theory, $\lambda_+$ and $\lambda_-$ are independent (real) couplings.

We will take the field $X$ to live on a circle of radius $R$, anticipating that the Lorentzian background of interest has thermal features. In order for the perturbation \eqref{defE} to make sense, the momentum $p$ must be an integer multiple of $1/R$. We will see that the discussion simplifies significantly if we take 
\begin{equation}
\label{defT}
p=\frac1R~,
\end{equation}
\IE the perturbation \eqref{defE} carries one unit of quantized momentum. We do not have a good understanding why this is the case, but will comment on what happens if we make other choices. 

We will use the Euclidean action \eqref{Sws}, \eqref{defE}, and the dual matrix model, to calculate some correlation functions of the form \eqref{vo},
\begin{equation}
\label{corE}
\big\langle \prod_{j=1}^nT_{q_j}\prod_{l=1}^{n'}T_{-q'_l}\big\rangle_{\lambda_+}
\end{equation}
(with $q_j,q'_l\ge0$). For $\lambda_+=0$, these correlation functions must satisfy the selection rule $\sum_jq_j=\sum_lq'_l$, due to $X$ translation invariance. However, for $\lambda_+>0$ this sum rule does not need to be satisfied, since the deformation \eqref{defE} breaks this symmetry. 

In particular, we will study the amplitudes \eqref{corE} with $n=0$, which vanish for $\lambda_+=0$, but don't for finite $\lambda_+$. We will also study the amplitudes with $n'=1$, for which the result can be read off from calculations in the existing literature.

We will then Wick rotate the Euclidean results \eqref{corE} to Lorentzian signature. We will find it convenient to first Fourier transform the momentum space results to position space, then Wick rotate the position space result, and finally Fourier transform back to (Lorentzian) momentum space, to obtain the S-matrix elements \eqref{corS}. 

The plan of the rest of the paper is the following. In section \ref{Euc}, we describe the procedure we employ in later sections for the case of an accelerating Liouville wall, for the usual static case. We start with Euclidean momentum space amplitudes, transform them to position space, then Wick rotate to Lorentzian signature, and finally Fourier transform back to Lorentzian momentum (energy) space. 

In section \ref{timed} we generalize this procedure to the time-dependent case. We consider backgrounds that correspond to a Liouville wall that is static in the far past, and approaches a finite velocity in the far future. The final velocity can be either smaller or larger than the speed of light, and we discuss both cases. For the former case we compute the scattering amplitude of $n$ incoming particles to one outgoing one (eq. \eqref{corS} with $n'=1$). We also compute the amplitude for creation of $n$ particles in the time-dependent background, \eqref{emit}. We show that these amplitudes have a thermal character, a phenomenon reminiscent of the Unruh effect~\cite{Unruh:1976db}. For the latter case, future null infinity is shielded by the potential, and we discuss the observables that one can define in this case.  

In section \ref{genz} we discuss some generalizations of the analysis of section \ref{timed}. The main goal of this section is to study the case where the Liouville wall accelerates both in the far past and in the far future, corresponding to eq. \eqref{defL} with $\lambda_\pm>0$ and figure \ref{3powa}. In the regime where the trajectory of the Liouville wall is timelike for all $t$, we compute the amplitude for particle creation. We find that these amplitudes have a singularity at a finite value of the coupling $\lambda_+\lambda_-$ \eqref{defL}. In a Hartle-Hawking construction, this singularity appears to correspond to the disappearance of time. 

In section \ref{fermis} we describe the results of the previous sections from the matrix model point of view. As is well known, in the double scaling limit the matrix model reduces to the dynamics of free fermions in an inverted quadratic potential. The massless tachyon field corresponds to a perturbation of the Fermi surface of these fermions. The time-dependent backgrounds we study in sections \ref{timed}, \ref{genz} correspond in this model to backgrounds with a time-dependent Fermi surface. We discuss the dynamics of perturbations of such surfaces, focusing on the difference between the cases $p<1$ and $p>1$. We relate the free fermion description to that in terms of an accelerating Liouville wall, and show that the dynamics of perturbations of the Fermi surface is compatible with that seen from the bulk $1+1$ dimensional string theory perspective. 

In section \ref{discuss} we  discuss our results and possible extensions. Two appendices contain reviews of technical results useful for our analysis.

\section{Wick rotation in the standard background}\label{Euc}

In this section we illustrate the procedure we will follow for the time-dependent backgrounds, in the usual, time-independent, background of two dimensional string theory, corresponding to a static Liouville wall. 

We start with the Euclidean theory \eqref{Sws} and take $X$ to live on a circle of radius $R$. In this theory, it is known from both continuum and matrix model perspectives \cite{Kutasov:1991pv,Ginsparg:1993is} that the Euclidean continuation of the $n\to 1$ scattering amplitude takes the form 
\ie
\label{TyTy}
\bigg\langle \prod_{j=1}^nT_{q_j}T_{-q'}\bigg\rangle= 2\pi R(-1)^n\partial_\mu^{n-2}\mu^{q'-1}\delta_{m',\ \sum_{j=1}^nm_j}\ .
\fe
Here $m'=q'R$ and $m_j=q_jR$, $j=1,\cdots, n$ are integer momenta. 

In order to Wick rotate this amplitude to Lorentzian signature, we proceed as follows. We start by Fourier transforming the operators $T_q$ \eqref{vo} to position space. We define a complex coordinate $y\equiv y_1+iy_2$, and write
\ie
\label{posspace}
T(y)=\frac{1}{2\pi R}\sum_{m=0}^\infty e^{-i\frac{m}{R}y}T_{\frac{m}{R}} ,\qquad \bar T(\bar{y})=\frac{1}{2\pi R}\sum_{m=0}^\infty e^{i\frac{m}{R}\bar{y}}T_{-\frac{m}{R}} \ .
\fe
As is clear from \eqref{posspace}, $y_1$ is a compact coordinate, $y_1\sim y_1+2\pi R$, so $y$ parameterizes a cylinder. 

Using \eqref{TyTy}, \eqref{posspace}, we find:
\ie
\label{poscor}
\bigg\langle \prod_{j=1}^nT(y_j)\bar{T}(\bar{y})\bigg\rangle 
= (2\pi R)^{-n}(-1)^n\partial_\mu^{n-2}\frac{1}{\mu}\prod_{j=1}^n\frac{1}{1-\mu^{\frac{1}{R}}e^{-\frac{i}{R}(y_j-\bar{y})}}\ .
\fe
We now Wick rotate $y_1=-iy_0$, which means $y=-iu$ and $\bar{y}=-iv$, where $u=y_0-y_2$, $v=y_0+y_2$ are lightlike coordinates on $R^{1,1}$. The corresponding Wick rotation in $(X,\phi)$ space is $X\to -it$, as described in section \ref{intro}. Looking back at \eqref{vo} we see that after Wick rotation one can identify $u$ with $t-\phi$ and $v$ with $t+\phi$, at least asymptotically (\IE near the boundary). 

After Wick rotation, the correlation function \eqref{poscor} takes the form 
\ie
\bigg\langle \prod_{j=1}^n\mathcal{T}^+(u_j)\mathcal{T}^-(v)\bigg\rangle= (2\pi R)^{-n}(-1)^n\partial_\mu^{n-2}\frac{1}{\mu}\prod_{j=1}^n\frac{1}{1-\mu^{\frac{1}{R}}e^{-\frac{1}{R}(u_j-v)}}~.
\fe
The next step is to Fourier transform back to Lorentzian momentum space. With
\ie
\label{Fourier}
\mathcal{T}^+_{\omega}=\int_{-\infty}^\infty du e^{-i\omega u}\mathcal{T}^+(u)~,\qquad \mathcal{T}^-_{\omega}=\int_{-\infty}^\infty dve^{i\omega v} \mathcal{T}^-(v)\ ,
\fe
the $n\to 1$ S-matrix elements are given by 
\ie
\label{corrnto1}
\bigg\langle \prod_{j=1}^n\mathcal{T}^+_{\omega_j}\mathcal{T}^-_{\omega'}\bigg\rangle= -(-2\pi)^{-n+1}\partial_\mu^{n-2}\frac{1}{\mu}\delta(\omega'-\sum_{j=1}^n\omega_j)\prod_{j=1}^n\mu^{-i\omega_j}g(\omega_jR)\ ,
\fe
where
\ie
\label{gw}
g(x)\equiv \int_{-\infty}^{+\infty} dz \frac{e^{ixz}}{1-e^z}\ .
\fe
The integral \eqref{gw} is divergent from $z\to -\infty$. This divergence can be regularized by an $i\epsilon$ prescription. Replacing $x\to x-i\epsilon$ in \eqref{gw}, the contribution to the integral from the region $z\to-\infty$ becomes convergent. 

The integrand in \eqref{gw} also has a pole at the origin of the $z$-plane, which can be treated as usual by deforming the integration contour to go either above or below it. In both cases the integral can be computed by standard contour deformation techniques. Closing the contour in the upper half plane, using the fact that in \eqref{corrnto1} we are interested in the case where $x$ in \eqref{gw} has a positive real part, we get that the integral receives contributions from the residues of the poles at $z=2\pi i k$, with integer $k$ ranging from 0 (or 1) to $\infty$:
\ie
\label{gwgw}
g(x)= -2\pi i \sum_{k=0\text{ or }1}^\infty e^{-2\pi x k}=-2\pi i\frac{e^{2\pi x}}{e^{2\pi x}-1}\;\text{ or }-2\pi i\frac{1}{e^{2\pi x}-1}~.
\fe
In our application \eqref{corrnto1}, $x=\omega R$, where $R$ is related to the temperature via $\beta=1/T=2\pi R$. At zero temperature, $R\to\infty$, with the first choice in \eqref{gwgw}, $g(\omega R)$ approaches a constant, while for the second it goes to zero. Therefore, to get a finite zero temperature limit, we choose the first expression in \eqref{gwgw}.
We have checked that the finite temperature amplitudes \eqref{corrnto1}, obtained via the Wick rotation described above, are consistent with those calculated in \cite{Balthazar:2022apu}.

In the limit $R\to\infty$, the amplitude \eqref{corrnto1} takes the form:
\ie
\label{zeroT}
\bigg\langle \prod_{j=1}^n\mathcal{T}^+_{\omega_j}\mathcal{T}^-_{\omega'}\bigg\rangle=2\pi i^n\delta(\omega'-\sum_{j=1}^n\omega_j)\partial_\mu^{n-2}\mu^{-i\sum_{j=1}^n\omega_j-1}\ .
\fe
This result is consistent with what we would have obtained by a simple continuation of \eqref{TyTy} to Lorentzian signature, but in the present discussion it is obtained as a limit from finite temperature of \eqref{corrnto1}, which contains more information. In particular, the factor $g(\omega R)$ for each incoming particle can be written as 
\ie
\label{gtt}
g=-2\pi i\frac1{1-e^{-\frac\omega T}}~.
\fe
This function is constant for $\omega\gg T$, but varies for $\omega\sim T$, and in fact diverges as $\omega\to 0$. 

In section \ref{intro} we discussed the fact that an insertion of a zero momentum tachyon can be thought of as differentiating the path integral w.r.t. $\mu$. We see that at finite temperature in Lorentzian signature, the situation is a bit more subtle. Equations \eqref{corrnto1}, \eqref{gtt} imply that 
\ie
\label{rescal}
\lim_{\omega\to 0}\frac{2\pi\mathcal{T}^+_{\omega}}{g(\omega R)}=\lim_{\omega\to 0}\frac{i\omega\mathcal{T}^+_{\omega}}{T}=-\partial_\mu~.
\fe
This is different from the situation at zero temperature, where we can see from \eqref{zeroT} that the limit $\omega\to 0$ of $\mathcal{T}^+_{\omega}$ gives $\partial_\mu$ times a multiplicative, $\omega$-independent factor. It is not surprising that the two cases are different, since the range of energies $\omega\ll T$ that is responsible for \eqref{rescal} does not exist for $T=0$.

\section{Amplitudes in time-dependent backgrounds}\label{timed}

In this section we generalize the discussion of section \ref{Euc} to the case of the time-dependent backgrounds described in section \ref{intro}. The starting point of that discussion was a Euclidean analysis at a finite value of the radius of Euclidean time, $R$, and it is natural to generalize it to the present case.

The calculation of section \ref{Euc} can be done for any $R$, but in the perturbed system \eqref{defE}, $R$ must satisfy the constraint $pR=m_p\in \mathbb{Z}$. Here $m_p$ is the quantized momentum in the Euclidean time direction. As mentioned in section \ref{intro}, the analysis simplifies significantly for $m_p=1$, \IE for $R=1/p$ as in \eqref{defT}, and we will mainly restrict to this case below. We will comment in the next section on what happens for $m_p>1$.

%Note that the background of figure \ref{2powa} has some features in common with reflection from a mirror that undergoes a period of acceleration \cite{birrell_davies_1982}, and therefore it is natural that the temperature is fixed in terms of $p$ via the relation above.

%We will focus on the case $m_p=1$, \IE for $R=1/p$ as in \eqref{defT}. 

%As mentioned in section \ref{intro}, the analysis simplifies significantly for $m_p=1$, \IE for $R=1/p$ \eqref{defT}, but  we do not have a good understanding why this is the right choice. However, we note that it is natural to expect that the background of figure \ref{2powa} has thermal properties. Indeed, it has some features in common with reflection from a mirror that undergoes a period of acceleration \cite{birrell_davies_1982}. It is natural that the temperature is fixed in terms of $p$, and \eqref{defT} is a natural choice. We will comment below on what happens when one takes $m_p>1$. \bruno{I think I counted the word natural 3 times in the last 3 sentences. I can reword this a bit if that's ok with everyone.} \david{Please do}

In the rest of this section we will study some examples of amplitudes in the background \eqref{defL} with $\lambda_-=0$. We start with the $n\to 1$ amplitude, the analog of \eqref{corrnto1} for non-zero $\lambda_+$, and then move on to the amplitude for production of $n$ particles. 

We also note that the calculations in this section involve the connected diagrams with given external legs. In general, these amplitudes receive contributions from disconnected diagrams as well, which are in fact generally dominant in the weak coupling limit. 

\subsection{$n\to 1$ amplitude}\label{ntoone}

To generalize the analysis of section \ref{Euc} to the case of non-zero $\lambda_+$, we need first to find the analog of \eqref{TyTy} for this case. In the Euclidean theory with compact $X$, we expect the Euclidean amplitude to be perturbative in $\lambda_+$. Expanding the exponential $\exp(-\lambda_+ T_p)$, we need to evaluate amplitudes of the form $\langle T_p^m\prod_{j=1}^nT_{q_j}T_{-q'}\rangle$ in the standard background of two dimensional string theory. The momenta $q_j, q'$ are integer multiples of $\frac{1}{R}$, 
\ie
\label{qqprime}
q_j=\frac{m_j}{R}~,\;\; q'=\frac{m'}{R}\ ,
\fe
as in \eqref{TyTy}. For $p=1/R$, \IE $m_p=1$, the momenta \eqref{qqprime} are integer multiples of $p$, but the analysis of this subsection can be performed for any $m_p$. As mentioned above, we will restrict to the case $m_p=1$ in this section, and will comment on $m_p>1$ later.  

The amplitude of interest is essentially the same as \eqref{TyTy}. It is given by 
\ie
\label{tachone}
\bigg\langle T_p^m\prod_{j=1}^nT_{q_j}T_{-q'}\bigg\rangle= \frac{2\pi}{p} (-1)^{m+n}\partial_\mu^{m+n-2}\mu^{q'-1}\delta_{m+\sum_{j=1}^nm_j, m'}~,
\fe
where the correlator on the l.h.s. is computed at $\lambda_+=0$, and $q_j$, $q'$ are given by \eqref{qqprime}. At finite $\lambda_+$, we have 
\ie
\label{mmm}
\bigg\langle \prod_{j=1}^nT_{q_j}T_{-q'}\bigg\rangle_{\lambda_+}=(-1)^{n}\frac{2\pi}{p}\sum_{m=0}^\infty \frac{\lambda_+^m}{m!}\partial_\mu^{m+n-2}\mu^{q'-1}\delta_{m+\sum_{j=1}^nm_j, m'}\ .
\fe
As in section \ref{Euc}, we next pass to position space using  \eqref{posspace}, which gives
\ie
\bigg\langle \prod_{j=1}^nT(y_j)\bar{T}(\bar{y})\bigg\rangle_{\lambda_+}=\left(-\frac{p}{2\pi}\right)^{n}
\sum_{m=0}^\infty e^{imp\bar{y}}\partial_\mu^{m+n-2}\mu^{mp-1}\frac{\lambda_+^m}{m!}
\prod_{j=1}^n\frac{1}{1-e^{-ip(y_j-\bar{y})}\mu^p}\ .
\fe
Next we Wick rotate the position space expression, by replacing $y_j\to -iu_j$, and $\bar y\to -iv$. We find
\ie
\bigg\langle \prod_{j=1}^n\mathcal{T}^+(u_j)\mathcal{T}^-(v)\bigg\rangle_{\lambda_+}=\left(-\frac{p}{2\pi}\right)^{n}\sum_{m=0}^\infty \frac{\lambda_+^m}{m!}e^{mpv} \partial_\mu^{m+n-2}\mu^{mp-1}
\prod_{j=1}^n\frac{1}{1-e^{-p(u_j-v)}\mu^p}\ .
\fe
Fourier transforming back to (Lorentzian) momentum space, as in \eqref{Fourier}, we find 
\ie
\label{nto1int}
\bigg\langle \prod_{j=1}^n\mathcal{T}^+_{\omega_j}\mathcal{T}^-_{\omega'}\bigg\rangle_{\lambda_+}=-(-2\pi )^{-n}\int_{-\infty}^{+\infty} dve^{i(\omega'-\sum_{j=1}^n\omega_j)v}\partial_\mu^{n-1}\mu^{-i\sum_{j=1}^n\omega_j}\frac{(1-r)^{i\sum_{j=1}^n\omega_j}}{i\sum_{j=1}^n\omega_j}\prod_{j=1}^ng(\omega_j/p)\ ,
\fe
where
\ie
\label{sv}
\frac{r}{(1-r)^{1-p}}\equiv e^{pv}\mu^{p-1}\lambda_+~, 
\fe
and $g(x)$ is given by \eqref{gwgw}.

For $p<1$ and $\lambda_+>0$, as $v$ varies from $-\infty$ to $\infty$, $r$ varies from 0 to 1. Changing variables from $v$ to $r$ in \eqref{nto1int}, we find
\ie
\label{nto1lambge0}
\bigg\langle \prod_{j=1}^n\mathcal{T}^+_{\omega_j}\mathcal{T}^-_{\omega'}\bigg\rangle_{\lambda_+}=-(-2\pi )^{-n}\frac{\Gamma(\alpha)\Gamma(\beta)}{p\Gamma(\alpha+\beta+1)}\partial_\mu^{n-1}\mu^{-\beta}\lambda_+^{-\alpha}\prod_{j=1}^ng(\omega_j/p)\ ,
\fe
where
\ie
\label{albeta}
\alpha= \frac{i}{p}(\omega'-\sum_{j=1}^n\omega_j)\equiv
\frac{i}{p}\Delta\omega~,
\qquad \beta= \frac{i}{p}\sum_{j=1}^n\omega_j+i\frac{p-1}{p}\omega'= i\omega'-\alpha~.
\fe
Thus, $\alpha$ measures the extent to which energy conservation is violated, and $\alpha+\beta$ measures the energy of the outgoing particle. 

We next discuss some special cases of the general expression \eqref{nto1lambge0}. When any of the incoming momenta $\omega_j$ goes to zero, we can use \eqref{rescal} to eliminate the corresponding operator from the correlation function. One can check that the structure of \eqref{nto1lambge0} is consistent with the resulting relation between the amplitudes with $n$ and $n-1$ incoming particles. 

Another interesting limit is the one in which the process \eqref{nto1lambge0} conserves energy, which corresponds to $\alpha\to 0$ in \eqref{nto1lambge0}. In this limit the amplitude approaches 
\ie
\label{scviol}
\bigg\langle \prod_{j=1}^n\mathcal{T}^+_{\omega_j}\mathcal{T}^-_{\omega'}\bigg\rangle_{\lambda_+}=(-2\pi)^{-n}\left(-\frac{\ln \lambda_+}{p}\right)\partial_\mu^{n-2}\mu^{-i\omega'-1}\prod_{j=1}^ng(\omega_j/p)~.
\fe
The structure of \eqref{scviol} can be understood as follows. An energy conserving process does not require the time-dependent perturbation proportional to $\lambda_+$ -- it can happen arbitrarily far from the time-dependent part of the Liouville wall. The $\ln \lambda_+$ dependence in \eqref{scviol} can be thought of as parametrizing the length of time up to the point \eqref{ptstarst}, available to the energy conserving process. Indeed, eq. \eqref{scviol} is the same as \eqref{corrnto1}, with the energy conservation $\delta$-function in the latter equation replaced by $-\frac{\ln \lambda_+}{p}$, as expected from the discussion above. 

For $n=0$, the correlation function \eqref{nto1lambge0} reduces to the one point function 
\ie
\label{1ptLo}
\big\langle \mathcal{T}^-_{\omega'}\big\rangle_{\lambda_+}=\frac{\Gamma(\frac{i}{p}\omega')\Gamma(i\frac{p-1}{p}\omega'-1)}{p\Gamma(i\omega'+1)}\mu^{-i\frac{p-1}{p}\omega'+1}\lambda_+^{-\frac{i}{p}\omega'}\ .
\fe
One can think of \eqref{1ptLo} as the amplitude for creation of one outgoing particle with energy $\omega'$ in the time-dependent background under consideration. We will discuss the (connected) amplitude for production of multiple particles in the next subsection. 

%Finally, above we analyzed the case $\lambda_+>0$. For negative $\lambda_+$ (but still $p<1$), the transform \eqref{sv} implies that as $v$ varies between $-\infty$ and $\infty$, $s$ varies between $-\infty$ and $0$. Repeating the analysis for this case gives 
%\ie
%\label{nto1lamble0}
%\bigg\langle \prod_{j=1}^n\mathcal{T}^+_{\omega_j}\mathcal{T}^-_{\omega'}\bigg\rangle_{\lambda_+}=(-2\pi )^{-n}\frac{\Gamma(\alpha)\Gamma(-i\omega')}{p\Gamma(1-\beta)}\partial_\mu^{n-1}\mu^{-\beta}|\lambda_+|^{-\alpha}\prod_{j=1}^ng(\omega_j/p)\ ,
%\fe
%with $\alpha$, $\beta$ as before, \eqref{albeta}. Comparing \eqref{nto1lamble0} to
%\eqref{nto1lambge0} we see that the ratio of the two is
%\ie
%-\frac{\sinh\pi\left(\omega'-\frac{\Delta\omega}{p}\right)}{\sinh\pi\omega'}\ ,
%\fe
%where $\Delta\omega$ is defined in eq. \eqref{albeta}.

\subsection{Particle production}\label{parprod}

In this subsection we compute the amplitudes for the production of $n$ particles of energies $(\omega_1,\omega_2,\cdots,\omega_n)$ in the time-dependent background \eqref{defL} with $\lambda_-=0$. These amplitudes were discussed in section \ref{intro}, around eq. \eqref{emit}, \eqref{KPZemit} -- \eqref{factorminus}. Here we will omit the primes on the various variables, for notational simplicity. These amplitudes are only defined for $p<1$, since for $p>1$ null future infinity is not part of the boundary. We will comment on the case $p>1$ later in this section. 

We follow the same procedure as in the previous subsection. Thus, our first goal is to compute the Euclidean amplitude $\langle T_{-q_1}T_{-q_2}\cdot\cdot\cdot T_{-q_n}\rangle_{\lambda_+}$. In order to do that, we need to compute the amplitudes 
$\langle T_p^mT_{-q_1}T_{-q_2}\cdot\cdot\cdot T_{-q_n}\rangle$ in the standard $c=1$ background. Unlike the situation in subsection \ref{ntoone}, here these amplitudes were not computed in the literature. Therefore, we calculated them ourselves, using the (matrix model) techniques of \cite{Moore:1991zv}. The calculation is described in appendix \ref{pp}.

The calculations simplify significantly for the case $m_p=1$, \IE $R=1/p$, \eqref{defT}. In this subsection we will present the results for this case. We found the following expression for the above correlation functions~\footnote{As a consistency check, if we take all $q_j=p$, \IE all $m_j=1$, \eqref{ttmmii} reduces to the amplitude computed in~\cite{Moore:1992ga}, and the results agree.}:
\ie
\label{ttmmii}
\langle T_p^mT_{-q_1}T_{-q_2}\cdot\cdot\cdot T_{-q_n}\rangle
=(-1)^{n-1}\frac{2\pi}{p}\frac{m!\partial_\mu^{n-3}\mu^{(p-1)m-1}\prod_{j=1}^n\Gamma\left((1-p)m_j+1\right)}{\prod_{j=1}^n\Gamma(m_j+1)\Gamma(1-pm_j)}\delta_{m,\sum_{j=1}^nm_j}\ .
\fe
Here $q_j$ are integer multiples of $p$, $q_j=m_jp$, as in \eqref{qqprime} with \eqref{defT}. 

Eq. \eqref{ttmmii} leads to 
\ie
\label{tmilamb}
\langle T_{-q_1}T_{-q_2}\cdot\cdot\cdot T_{-q_n}\rangle_{\lambda_+}
=(-1)^{n-1}\frac{2\pi}{p}\frac{(-\lambda_+)^{\sum_{j=1}^nm_j}\partial_\mu^{n-3}\mu^{(p-1)\sum_{j=1}^nm_j-1}\prod_{j=1}^n\Gamma\left((1-p)m_j+1\right)}{\prod_{j=1}^n\Gamma(m_j+1)\Gamma(1-pm_j)}\ ,
\fe
and in position space \eqref{posspace},
\ie
\label{corry1}
\langle \bar{T}(\bar{y}_1) \bar{T}(\bar{y}_2)\cdot\cdot\cdot \bar{T}(\bar{y}_n)\rangle_{\lambda_+}
=\left(-\frac{p}{2\pi}\right)^{n-1}\partial_\mu^{n-3}\frac{1}{\mu}\prod_{j=1}^n f(e^{ip\bar{y}_j}\lambda_+\mu^{p-1})\ ,
\fe
where~\cite{alma991033303509703276}
\ie
\label{deffx}
f(x)\equiv &\sum_{m=0}^\infty \frac{(-x)^m\Gamma\left(m(1-p)+1\right)}{m!\Gamma(-mp+1)}\\
=&-\frac{\partial}{\partial x}\sum_{m=0}^\infty \frac{(-x)^m\Gamma\left(m(1-p)+p\right)}{m!\Gamma(-mp+p+1)}\\
=&-\frac{\partial}{\partial x}\frac{(1-r)^p}{p}\ ,
\fe
and $r(x)$ is defined similarly to \eqref{sv},
\ie
\label{defsss}
\frac{r}{(1-r)^{1-p}}= x\ .
\fe
Wick rotating as before, $\bar{y}_j=-iv_j$, gives
\ie
\label{positionspace}
\langle \mathcal{T}^-(v_1) \mathcal{T}^-(v_2)\cdot\cdot\cdot \mathcal{T}^-(v_n)\rangle_{\lambda_+}
=\left(-\frac{p}{2\pi}\right)^{n-1} \partial_\mu^{n-3}\frac{1}{\mu}\prod_{j=1}^n f(e^{pv_j}\lambda_+\mu^{p-1})\ .
\fe
Finally, Fourier transforming to $\omega$ space, as in \eqref{Fourier}, we find 
\ie
\label{finminus}
\langle \mathcal{T}^-_{\omega_1} \mathcal{T}^-_{\omega_2}\cdot\cdot\cdot \mathcal{T}^-_{\omega_n}\rangle_{\lambda_+}
=\left(-\frac{1}{2\pi}\right)^{n-1}\frac{1}{p}\partial_\mu^{n-3}\frac{1}{\mu}\prod_{j=1}^n\frac{\Gamma\left(\frac{i\omega_j}{p}\right)\Gamma\left(\frac{p-1}{p}i\omega_j+1\right)}{\Gamma(i\omega_j+1)}(\lambda_+\mu^{p-1})^{-i\frac{\omega_j}{p}}\ .
\fe
To better understand the structure of \eqref{finminus}, it is useful to consider the position space correlator \eqref{positionspace}. The basic building block of this correlator is a factor of the function $f(x)$ \eqref{deffx}, with
\ie
\label{xv}
x=e^{pv}\lambda_+\mu^{p-1}~, 
\fe
for each external leg. The momentum space correlator \eqref{finminus} is obtained from $f$ by Fourier transforming in $v$, whch is related to $x$ via \eqref{xv}. 

\begin{figure}[h!]
\centering
 {\subfloat{\includegraphics[width=0.4\textwidth]{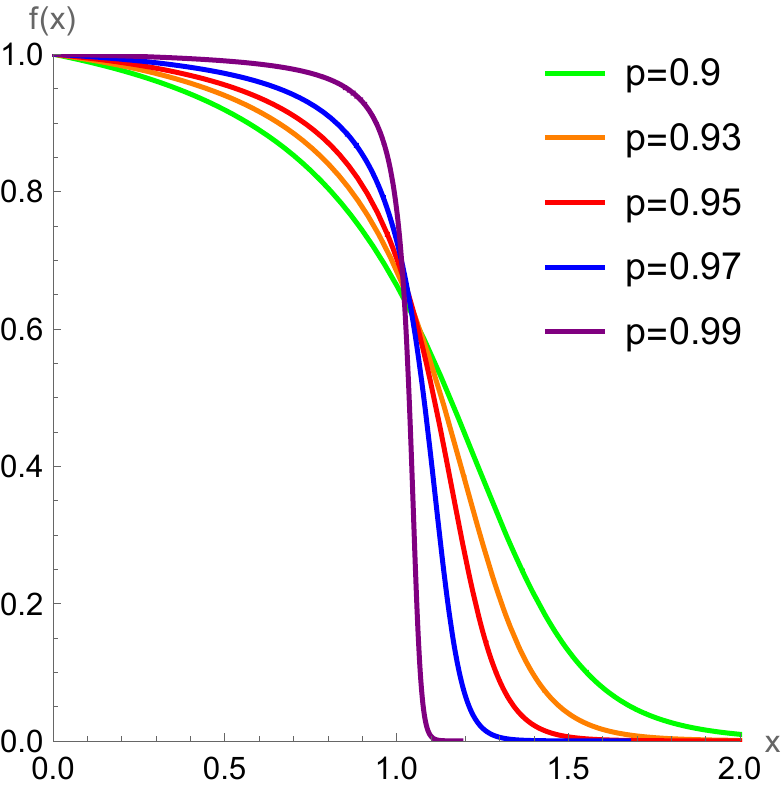}\label{fx099}}}~ 
  \hspace{1.5 cm}
 {\subfloat{\includegraphics[width=0.4\textwidth]{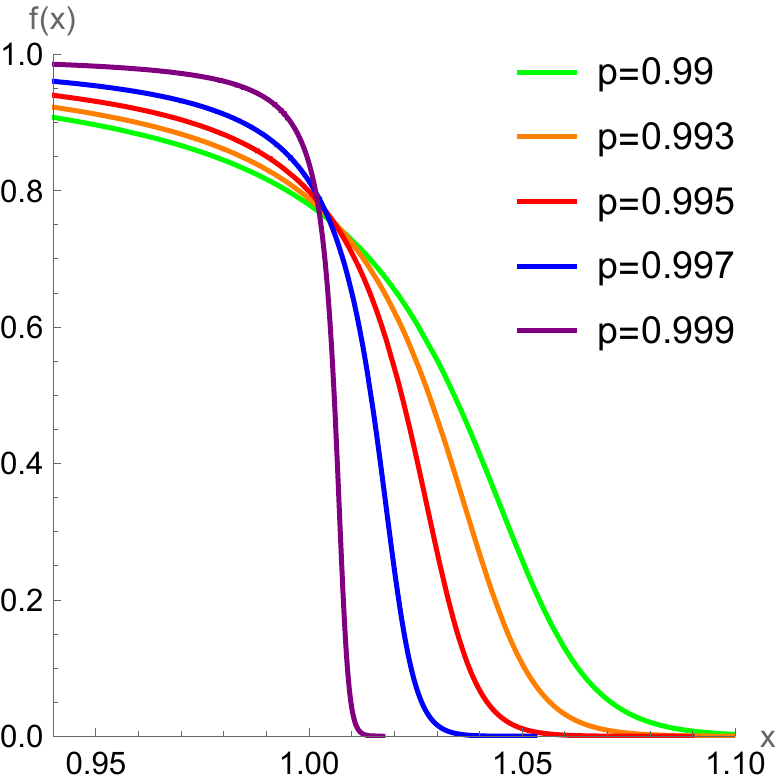}\label{fx0999}}} 
 \caption{The position space amplitude \eqref{positionspace} includes a factor of the function $f$ for each created particle. In this figure we plot this function for a range of values of $p$, and demonstrate that it approaches a step function as $p\to 1$.}
 \label{fx}
\end{figure}

In the left panel of figure \ref{fx} we plot the function $f(x)$ for a few values of $p$. We chose values close to one since, as we discussed in section \ref{intro}, one of the interesting issues is what happens to the particle creation amplitudes as $p\to 1$, a point beyond which future null infinity disappears. We see that as $p$ approaches $1$, the function $f(x)$ approaches a step function centered at $x=1$. In the right panel of figure \ref{fx} we verify this by focusing on the behavior of the function $f$ very close to $x=1$ for $p$ very close to one. 

The form of the function $f$ in figure \ref{fx} seems in agreement with the physical picture suggested by figure \ref{2powa}. The particles are created throughout the whole process of the acceleration of the Liouville wall, which corresponds to $v$ between $-\infty$ and $+\infty$, and indeed for general $p$ between $0$ and $1$ the function $f$ has support for all $x$ in the range $(0,\infty)$. However, as $p$ approaches $1$, the velocity of the wall at late times approaches the speed of light, and {\it at} $p=1$ there is an upper bound on the value of $v$, corresponding to the position of the wall. Thus, it is natural that in this limit $f$ approaches a step function. The location of the step is at $v^*=t^*+\phi^*$, where $(\phi^*,t^*)$ appear in figure \ref{2powa} and are given by eq. \eqref{ptstarst}. One can check that eq. \eqref{ptstarst} implies that $v^*=t^*+\phi^*$ is equal to the $v$ that corresponds to $x=1$ in figure \ref{fx} via eq. \eqref{xv}. 

The steep decline of the function $f$ near $x=1$ for $p$ close to $1$ corresponds physically to the fact that as $p\to 1$ the region $v\simeq v^*$ corresponds to larger and larger $u(=t-\phi)$, where the speed of the Liouville wall is already approximately equal to its final value. Thus, this region doesn't radiate, and its contribution to the Fourier transform performed in going from \eqref{positionspace} to \eqref{finminus} is small. 

It is interesting to take the limit $\omega\to 0$ of \eqref{finminus}, following the same logic as in \eqref{rescal}. To get a finite answer, we must take the limit as follows:
\ie
\label{limminus}
\lim_{\omega\to 0}\frac{2\pi i\omega\mathcal{T}^-_{\omega}}{p}=-\partial_\mu\ .
\fe
Recalling that the temperature $T$ is given by $T=1/2\pi R=p/2\pi$, we see that the limit \eqref{limminus} is compatible with that taken in \eqref{rescal}. 

Another interesting limit is $p\to 0$. As is clear from eq. \eqref{wspot} and figure \ref{2powa}, in this limit we go back to the time-independent background corresponding to a stationary Liouville wall of section \ref{Euc}. Thus, in this limit the particle production amplitudes \eqref{finminus} should go to zero. To see that this is indeed the case, note that in this limit the phase in \eqref{finminus}, which is given in \eqref{factorminus}, together with a contribution from the beta function prefactor, is more and more rapidly oscillating for any non-zero $\omega$. Thus, the amplitude to create  particles with a finite distribution of energies goes to zero in the limit $p\to 0$.

The above discussion also has an interesting consequence for the results of subsection \ref{ntoone}. Equation \eqref{nto1lambge0} has a finite limit as $\omega'\to 0$:
\ie
\label{ttone}
\bigg\langle \prod_{j=1}^n\mathcal{T}^+_{\omega_j}\mathcal{T}^-_0\bigg\rangle_{\lambda_+}=\frac{(-1)^{n+1}(2\pi)^{-n}\pi}{\sum_{j=1}^n\omega_j}\sinh^{-1}\left(\frac{\pi\sum_{j=1}^n\omega_j}{p}\right)\partial_\mu^{n-1}\mu^{-\frac{i}{p}\sum_{j=1}^n\omega_j}\lambda_+^{\frac{i}{p}\sum_{j=1}^n\omega_j}\prod_{j=1}^ng(\omega_jR)\ .
\fe
Using \eqref{limminus}, we conclude that the amplitude \eqref{KPZabsor} vanishes, for all $n$ and energies $\{\omega_j\}$. In other words, the amplitude for a process in which $n$ tachyons come in from past null infinity, are absorbed by the time-dependent Liouville wall, and no tachyons come out at late time, vanishes for all $p<1$.

\subsection{$p>1$ }\label{pgone}

The general discussion in section \ref{intro} leads one to believe that for $p>1$ the observables $\mathcal{T}_\omega^-$ \eqref{voL} should be problematic. Indeed, looking back at figure \ref{2powa}, we see that for $p>1$ future null infinity is shielded by the dynamical Liouville wall, so an S-matrix cannot be defined. It is interesting to see how this is reflected in the calculations reported earlier in this section. This is the goal of this subsection. 

Following the discussion of subsection \ref{parprod}, the Euclidean results \eqref{ttmmii} -- \eqref{corry1} are clearly valid for all $p$. We can again Wick rotate and get the position space correlation function \eqref{positionspace}. An important difference between the case $p<1$ and the present one concerns equation \eqref{defsss}. Recalling that the variable $x$ in that equation is related to $v$ via eq. \eqref{xv}, we see that for $p<1$ as $v$ goes from $-\infty$ to $\infty$, $x$ goes from $0$ to $\infty$, and $r$ goes from $0$ to $1$. However, for $p>1$ the situation is different. In that case, one can write \eqref{defsss} as $x=r(1-r)^{p-1}$ and as $r$ goes from $0$ to $1$, $x$ first increases, from $0$ to a maximal value, $x_{\rm max}$, and then decreases back to $0$ at $r=1$. The maximum of the function $x(r)$ occurs at $r=1/p$, so 
\ie
\label{xxmax}
x_{\rm max}=x(1/p)=(p-1)^{p-1}p^{-p}~.
\fe
In terms of the variable $v$ \eqref{xv}, as $r$ varies between $0$ and $1$, $v$ increases from $-\infty$, to some $v_{\rm max}$, related to $x_{\rm max}$ \eqref{xxmax} via \eqref{xv}, and then goes back to $-\infty$. This means that the Fourier transform \eqref{Fourier} that was done for $p<1$ in going from position space, \eqref{positionspace}, to momentum space, \eqref{finminus}, can no longer be done since the range of $v$ does not extend all the way to $+\infty$. 

The above discussion is in nice correspondence with properties of the potential of figure \ref{2powa} and eq. \eqref{sppot}. The solid line in that figure is obtained by setting $V_{\rm st}(t,\phi)=1$. Since we are interested in the process of emission of tachyons from the accelerating Liouville wall, we are interested in the value of $v$ along the wall. It is clear from figure \ref{2powa1}, that for $p<1$, as we move along the Liouville wall, $v$ varies from $-\infty$ to $\infty$, but for $p>1$, the behavior is different. 

In that case, the wall starts at $v=-\infty$ at early times, moves to larger $v$, but then reaches a maximal value of $v$, $v_{\rm max}$ in figure \ref{2powa2}, at which its velocity reaches the speed of light, and then it starts decreasing, going back to $v=-\infty$ at late times. A short calculation using eq. \eqref{sppot} shows that
\ie
\label{valvmax}
v_{\rm max}=\frac{1}{p}\ln \frac{(p-1)^{p-1}}{p^p\lambda_+\mu^{p-1}}\ .
\fe
In our discussion of the function $x(r)$ above, we obtained the value of $v_{\rm max}$ by plugging \eqref{xxmax} into \eqref{xv}. Interestingly, the two values of $v_{\rm max}$ computed in these different ways coincide.\footnote{In general, we expect agreement up to an order one constant, that has to do with the fact that we could have set $V_{\rm st}(t,\phi)$ \eqref{sppot} to a  constant $\not=1$, but as it happens, for $V=1$ the agreement appears to be exact.}

 This also explains why the function $x(r)$ has the property (for $p>1$) that each $x<x_{\rm max}$ appears twice as $r$ varies from $0$ to $1$. In terms of figure \ref{2powa2}, this is a reflection of the fact that each $v<v_{\rm max}$ appears twice, once in the timelike part of the trajectory of the Liouville wall, and once in the spacelike one. The former corresponds to $0<r<1/p$; the latter, to $1/p<r<1$.

\begin{figure}[h]
\centering
 \includegraphics[width=0.5\textwidth]{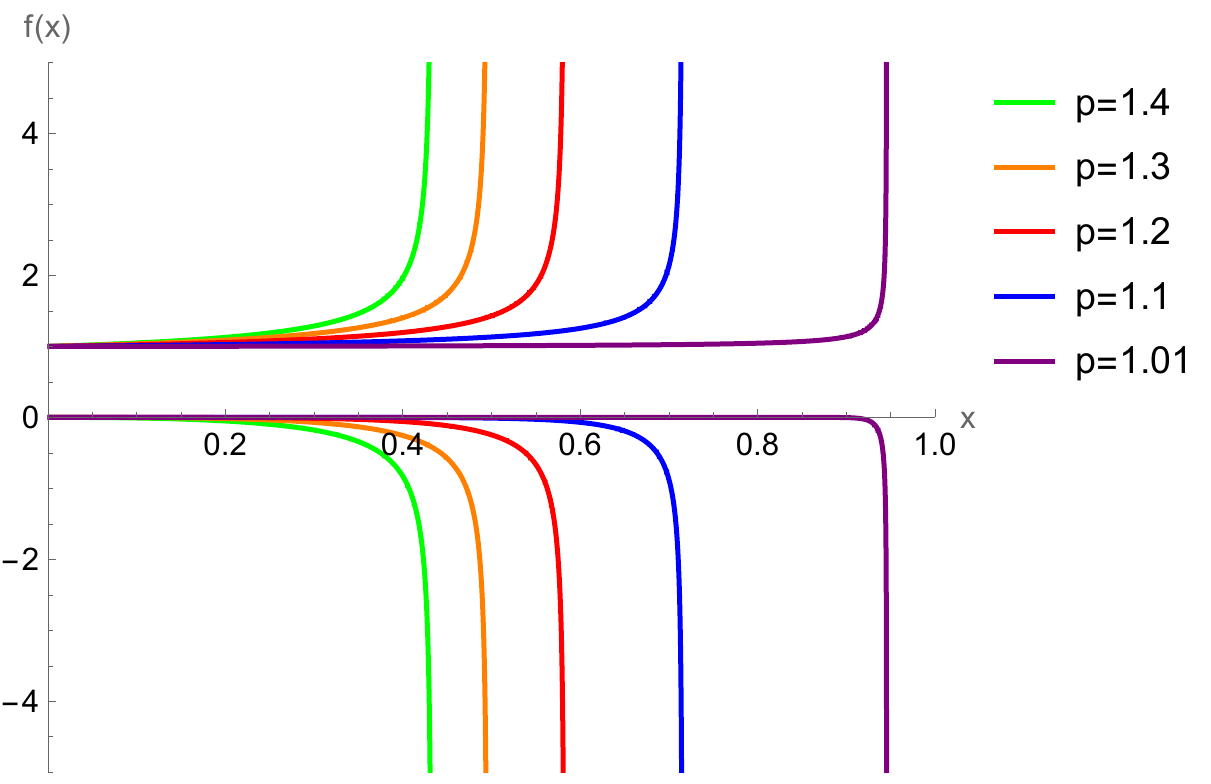}
 \caption{The form of the function $f(x)$, \eqref{deffx}, for $p>1$. The singularity of the function is at $x=x_{\rm max}$ \eqref{xxmax}. The two branches correspond to $r<1/p$ (upper) and $r>1/p$ (lower) or, equivalently, to the timelike and spacelike parts of the trajectory of the Liouville wall.} 
\label{fxpg1}
\end{figure}

As for $p<1$, it is interesting to plot the function $f(x)$ for $p>1$. We do this for a few values of $p$ in figure \ref{fxpg1}. As expected, the function $f$ has two branches. The upper one corresponds to the timelike part of the trajectory of the Liouville wall. It starts at $x=0$ (\IE $r=0$, $v=-\infty$) at $f=1$, just like for $p<1$ (figure  \ref{fx}), but in this case it monotonically increases, and diverges as $x\to x_{\rm max}$, \eqref{xxmax}. This divergence appears to be due to the fact that the acceleration of the Liouville wall diverges as its speed approaches that of light. 

Viewed as a function of $r$, \eqref{defsss}, $f$ has a single pole at $r=1/p$. Thus, as we transition to the spacelike part of the trajectory of the Liouville wall, $r>1/p$, described by the lower branch of the curve in figure \ref{fxpg1}, $f$ starts from $-\infty$ and rapidly goes to zero. Its decline becomes more and more pronounced as $p\to 1$ (from above). 

In order to understand this behavior, one needs to make sense of the observables \eqref{positionspace} in this case. For $p<1$, they were related by a Fourier transform to amplitudes for detecting particles at future null infinity, \eqref{finminus}. As mentioned above, for $p>1$ such amplitudes do not make sense. This is reflected in our analysis in the fact that the variables $v_j$ in \eqref{positionspace} are in this case bounded from above by $v_{\rm max}$ \eqref{valvmax}. Thus, the usual Fourier transform cannot be done.

There are two attitudes that one can take to this state of affairs. One is that for $p>1$ there are no physical observables associated with the future region. This seems problematic, especially given the fact that as discussed in the introduction and in the next section, one can turn on both $\lambda_+$ and $\lambda_-$, \eqref{defL}, in which case all asymptotic particle states cease to exist, since both past and future null infinities are shielded by the potential. The other possible attitude is that the correlation functions \eqref{positionspace} do make sense for $p>1$, as seems to be the case from the above analysis. In that case, one needs to interpret them physically. We will return to this question in later sections.

\section{Generalizations}
\label{genz}

The main goal of this section is to discuss the generalization of the analysis of section \ref{timed} to the case where both $\lambda_+$ and $\lambda_-$ in \eqref{defL} are positive.  However, we start with two brief comments about other issues that were mentioned earlier. 

The first involves quantum corrections to the results of section \ref{timed}. The analysis of that section was done at leading order in $g_s$, \IE the worldsheet was taken to have spherical topology. The matrix model allows one to compute higher order (in $g_s$) corrections to these results. For example, in \eqref{TqTpmtorus} and \eqref{Tq1q2Tpmtorus} we present the results for the first subleading corrections to the Euclidean correlation functions $\langle T_0T_p^mT_{-q}\rangle$ and $\langle T_0T_p^mT_{-q_1}T_{-q_2}\rangle$, respectively. These corrections come from the worldsheet torus and, as in section \ref{timed}, they can be used to calculate the torus contributions to the one and two-point functions of $\mathcal{T}^-_\omega$. We will leave a detailed study of these contributions to future work. 

The second issue we want to comment on involves eq. \eqref{defT}. We said there (and in section \ref{timed}) that in the Euclidean calculation one could in principle consider other values of the quantized momentum, \IE take $pR=m_p>1$, but it seems that the right value is $m_p=1$. One of the reasons we believe this is that for $m_p>1$ the structure of the amplitudes becomes much more involved. Here we would like to explain what we mean by this. 

For general integer $m_p\ge 1$, the calculation that for $m_p=1$ leads to \eqref{tmilamb}, gives the following results. 
For $n=3$, $q_j=\frac{m_j}{R}$ and $\sum_{j=1}^nm_j$ an integer  multiple of $m_p$, we find
\ie
\label{n123}
\langle T_{-q_1}T_{-q_2} T_{-q_3}\rangle_{\lambda_+}
=\frac{2\pi R}{\mu}(\lambda_+\mu^{p-1})^{\frac{1}{m_p}\sum_{j=1}^3m_j}\prod_{j=1}^3h_1(q_j)~,
\fe
where
\ie
\label{formhone}
h_1(q)\equiv (p-1)^{q/p-[q/p]}\prod_{i=1}^{[q/p]}\left(\frac{q}{i}-1\right)~.
\fe
One can check that for $m_p=1$, \eqref{n123} and \eqref{formhone} agree with \eqref{tmilamb}. For $m_p>1$, \eqref{n123} is more complicated, but it follows the same pattern as \eqref{tmilamb} -- it is still a product of factors associated with the external legs. 

For the four point function the factorized structure breaks down. We find
\ie
\label{n4}
\langle T_{-q_1}T_{-q_2}T_{-q_3}T_{-q_4}\rangle_{\lambda_+}=-\frac{2\pi R}{\mu^2}(\lambda_+\mu^{p-1})^{\frac{1}{m_p}\sum_{j=1}^3m_j}
h_2(q_1,\cdots,q_4)\prod_{j=1}^4h_1(q_j)~,
\fe
where
\ie
\label{g1}
h_2(q_1,q_2,q_3,q_4)\equiv mp-m-1+\sum_{ T\subseteq \{q_1,q_2,q_3,q_4\}}\frac{(-1)^{|T|}\sum_{l=1}^{m-1}|q(T)-lp|}{4}\ .
\fe
Here $T$ is any subset of $ \{q_1,q_2,q_3,q_4\}$, and $|T|$ and $q(T)$ denote the number and sum of the elements in $T$, respectively. One can check that when $q_j/p$ are integers, which is the case for $m_p=1$, eq. \eqref{n4} reduces to \eqref{tmilamb}. However, for general $m_p>1$, $q_j/p$ is not an integer, and \eqref{n4} is more complicated. 

This complexity increases as the number of insertions $n$ increases, and we believe that the resulting theory does not correspond after Wick rotation to the theory described in section \ref{intro}. It would be interesting to understand whether it has a physical interpretation. This too is left for future work.  

\begin{figure}[h]
\centering
 \includegraphics[width=0.45\textwidth]{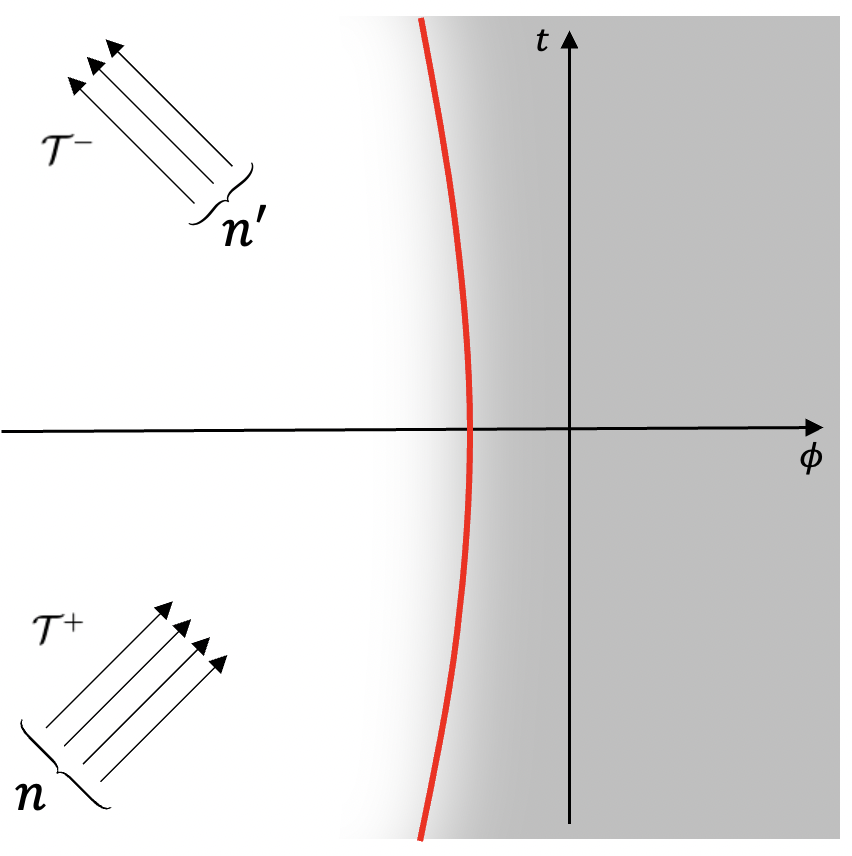}
 \caption{For general $\lambda_+$ and $\lambda_-$, the trajectory of the Liouville wall takes the qualitative form depicted here.
 } 
\label{3powa}
\end{figure}

We next turn to the case where both $\lambda_+$ and $\lambda_-$ in \eqref{defL} are positive. We start with a few comments:
\begin{itemize}
    \item In section \ref{timed} we discussed the case $\lambda_-=0$, in which the Liouville wall is stationary in the past and has a time-dependent form in the future. The case $\lambda_+=0$ can be similarly studied by taking $t\to -t$, and exchanging the roles of $\mathcal{T}^+$ and $\mathcal{T}^-$.
    \item Turning on both $T_p$ and $T_{-p}$ in \eqref{defL}, the potential \eqref{sppot} takes the form  
\begin{equation}
\label{sppotpm}
V_{\rm st}(t,\phi)=\mu e^{2 \phi}+\lambda_+e^{(2-p)\phi+pt}+\lambda_-e^{(2-p)\phi-pt}~.
\end{equation}    
We will take the couplings  $\lambda_\pm$ to be positive, as before. 
\item In principle, we could take the momentum of the $\lambda_-$ deformation in \eqref{sppotpm} to be different from (the negative of) that of the $\lambda_+$ one. However, as discussed above, our Euclidean space techniques make it natural to take both of them to be $1/R$, \eqref{defT}. This is consistent with  \cite{Alexandrov:2002pz}, who show that the perturbation \eqref{sppotpm} leads to thermodynamic behavior with temperature $p/2\pi$. Thus, we will restrict to the case where the two momenta are equal (and opposite).
\item When $\lambda_\pm>0$, we can set $\lambda_+=\lambda_-=\lambda$ in \eqref{sppotpm}, by shifting the origin of time. In this case, which is depicted in figure \ref{3powa}, the worldsheet theory has a symmetry $t\to -t$. This allows one to use the Hartle-Hawking construction: replace the $t<0$ region of the background by its Euclidean analog, \eqref{defE}, which similarly has an $X\to -X$ symmetry, and view it as providing an initial state for the Lorentzian evolution at $t>0$.
\item When $\lambda_\pm$ are both non-zero, there is an ambiguity in the definition of $\mu$. If either of these couplings vanishes, we can go to a region where the Liouville wall is stationary and read off $\mu$ there. However, when $\lambda_+\lambda_-\not=0$, such a region does not exist. This ambiguity will play a role below. 
\end{itemize}

In the rest of this section we will generalize the discussion of section \ref{timed} to the case $\lambda_+,\lambda_->0$. We will use the same method as there: start with the Euclidean calculation, go to position space, Wick rotate, and go back to (Lorentzian) momentum space. We will only consider the amplitude for emission of $n$ outgoing tachyons \eqref{emit}. The amplitude for absorption of $n$ tachyons, \eqref{absor}, can be obtained from it by using the time reversal symmetry mentioned above.  

Fortunately, the Euclidean amplitudes needed for the analysis have been calculated before. In particular, from \eqref{ttmmii}, we have
\ie
\langle T_{p}^{m+m'}T_{-p}^{m'}T_{-q_1}T_{-q_2}\cdots T_{-q_n}\rangle=&(-1)^{n+m'-1}\frac{2\pi}{p}(m+m')!\partial_\mu^{n+m'-3}\mu^{(p-1)(m+m')-1}\\
&(1-p)^{m'}\prod_{j=1}^n\frac{\Gamma\left((1-p)m_j+1\right)}{\Gamma(m_j+1)\Gamma(1-pm_j)}\delta_{m,\sum_{j=1}^nm_j}\ ,
\fe
and therefore
\ie
\langle T_{-q_1}T_{-q_2}\cdots T_{-q_n}\rangle_{\lambda_+,\lambda_-}=&\sum_{m,m'=0}^\infty\frac{(-\lambda_+)^{m+m'}(-\lambda_-)^{m'}}{(m+m')!m'!}\langle T_{p}^{m+m'}T_{-p}^{m'}T_{-q_1}T_{-q_2}\cdots T_{-q_n}\rangle\\
=&\sum_{m'=0}^\infty\frac{(-\lambda_+)^{\sum_{j=1}^nm_j+m'}(-\lambda_-)^{m'}}{m'!}(-1)^{n+m'-1}\frac{2\pi}{p}\partial_\mu^{n+m'-3}\\
&\mu^{(p-1)(\sum_{j=1}^nm_j+m')-1}(1-p)^{m'}\prod_{j=1}^n\frac{\Gamma\left((1-p)m_j+1\right)}{\Gamma(m_j+1)\Gamma(1-pm_j)}\ ,
\fe

Fourier transforming, as in \eqref{posspace}, \eqref{corry1}, gives
\ie
\langle \bar{T}(\bar{y}_1) \bar{T}(\bar{y}_2)\cdot\cdot\cdot \bar{T}(\bar{y}_n)\rangle_{\lambda_+,\lambda_-}
=&\sum_{m'=0}^\infty\frac{((p-1)\lambda_+\lambda_-)^{m'}}{m'!}\left(-\frac{p}{2\pi}\right)^{n-1}\partial_\mu^{n+m'-3}\\
&\mu^{(p-1)m'-1}\prod_{j=1}^n f(e^{ip\bar{y}_j}\lambda_+\mu^{p-1})\ ,
\fe
and Wick rotating, $\bar{y}_j=-iv_j$, we find
\ie
\label{corr+-v}
\langle \mathcal{T}^-(v_1) \mathcal{T}^-(v_2)\cdot\cdot\cdot \mathcal{T}^-(v_n)\rangle_{\lambda_+,\lambda_-}
=&\sum_{m'=0}^\infty\frac{((p-1)\lambda_+\lambda_-)^{m'}}{m'!}\left(-\frac{p}{2\pi}\right)^{n-1} \partial_\mu^{n+m'-3}\\
&\mu^{(p-1)m'-1}\prod_{j=1}^n f(e^{pv_j}\lambda_+\mu^{p-1})\ .
\fe

For $p<1$, we expect to be able to Fourier transform \eqref{corr+-v} to momentum space. Performing the transform, we find 
\ie
\langle \mathcal{T}^-_{\omega_1} \mathcal{T}^-_{\omega_2}\cdot\cdot\cdot \mathcal{T}^-_{\omega_n}\rangle_{\lambda_+,\lambda_-}
=&\sum_{m'=0}^\infty\frac{((p-1)\lambda_+\lambda_-)^{m'}}{m'!}\left(-\frac{1}{2\pi}\right)^{n-1}\frac{1}{p}\partial_\mu^{n+m'-3}\\
&\mu^{(p-1)m'-1}\prod_{j=1}^n\frac{\Gamma\left(\frac{i\omega_j}{p}\right)\Gamma\left(\frac{p-1}{p}i\omega_j+1\right)}{\Gamma(i\omega_j+1)}(\lambda_+\mu^{p-1})^{-i\frac{\omega_j}{p}}\ ,
\fe
and after summing over $m'$: 
\ie
\label{corr+-}
\langle \mathcal{T}^-_{\omega_1} \mathcal{T}^-_{\omega_2}\cdot\cdot\cdot \mathcal{T}^-_{\omega_n}\rangle_{\lambda_+,\lambda_-}
=&\left(-\frac{1}{2\pi}\right)^{n-1}\frac{1}{p}\partial_\mu^{n-3}\frac{1}{\mu}\frac{(1-s)^{i\frac{p-1}{p}\sum_{j=1}^n\omega_j+1}}{1+(1-p)s}\\
&\prod_{j=1}^n\frac{\Gamma\left(\frac{i\omega_j}{p}\right)\Gamma\left(\frac{p-1}{p}i\omega_j+1\right)}{\Gamma(i\omega_j+1)}(\lambda_+\mu^{p-1})^{-i\frac{\omega_j}{p}}\ ,
\fe
where
\ie
\label{mapsprime}
\frac{s}{(1-s)^{2-p}}=(p-1)\lambda_+\lambda_-\mu^{p-2}\ .
\fe
Note that for $p<1$, for given values of $\mu$ and $\lambda_\pm$, \eqref{mapsprime} has two solutions for $s$. One should take the one that goes to zero when $\lambda_-\to 0$, so that \eqref{corr+-} reduces to \eqref{finminus} in the limit. 

For general $s$, one can Fourier transform back to position space, and obtain the analog of \eqref{positionspace} for this case. One finds\footnote{This is a more explicit form of \eqref{corr+-v}.} 
\ie
\label{corr-fin}
\langle \mathcal{T}^-(v_1) \mathcal{T}^-(v_2)\cdot\cdot\cdot \mathcal{T}^-(v_n)\rangle_{\lambda_+,\lambda_-}
=\left(-\frac{p}{2\pi}\right)^{n-1} \partial_\mu^{n-3}\frac{1}{\mu}\frac{1-s}{1+(1-p)s}\prod_{j=1}^n f(e^{pv_j}(1-s)^{1-p}\lambda_+\mu^{p-1})\ .
\fe
We next comment on some properties of \eqref{corr+-} -- \eqref{corr-fin}.

One property is that for $p<1$,\footnote{For $p>1$ there is no such singularity, since $s=\frac{1}{p-1}$ is outside the physical regime $0\le s<1$.} \eqref{corr+-}, \eqref{corr-fin} diverge as $s\to\frac{1}{p-1}$. That value of $s$ corresponds to the maximal value of the coupling $\lambda_+\lambda_-$, according to \eqref{mapsprime}. The physics of this singularity was discussed in \cite{Hsu:1992cm,Kazakov:2000pm} in the Euclidean system. In that case, the worldsheet potential is proportional to $\cos pX$, and the singularity is associated with the field $X$ settling at the minimum of the potential and disappearing from the dynamics. 

In our Lorentzian system, this would correspond to the disappearance of time -- an exotic phenomenon. We leave a more complete understanding of this phenomenon to future work. Here, we note that a setting in which the question what happens as $s$ approaches the critical value $\frac{1}{p-1}$ can be addressed is the Hartle-Hawking construction mentioned above. 

Another interesting feature of \eqref{corr-fin} is that the effect of non-zero $s$ on the function $f$ is to change its argument by the multiplicative factor $(1-s)^{1-p}$. One possible way to interpret this is to use an observation we made earlier. When both couplings $\lambda_\pm$ are non-zero, the notion of what is $\mu$ is ambiguous. Thus, we can view the multiplicative factor in the argument of $f$ in \eqref{corr-fin} as due to a renormalization of $\mu$. Indeed, if we make the replacement
\ie
\label{renmu}
\mu\to\mu (1-s)~,
\fe
the argument of $f$ in \eqref{corr-fin} goes back to that in \eqref{positionspace}. This choice is natural since it also simplifies eq. \eqref{mapsprime}, which becomes\footnote{Note that since for $p>1$ \eqref{mapsprime} implies that $0\le s<1$, in terms of the rescaled $\mu$ the r.h.s. of \eqref{newmaps} is bounded from above. This is just a different parametrization of coupling space, a standard phenomenon in QFT.} 
\ie
\label{newmaps}
s=(p-1)\lambda_+\lambda_-\mu^{p-2}\ .
\fe
Note also that the replacement \eqref{renmu} was found in \cite{Kazakov:2000pm} to simplify the expression for the partition sum of the Euclidean theory \eqref{defE}; see the discussion around eq. (3.11) in that paper.

We will see that the replacement \eqref{renmu} is also natural from the point of view of the dynamics of the free fermions in the matrix model. However, whether we make this replacement or not, the important thing is that for non-zero $s$ we have an ambiguity of rescaling  $\mu$ by a function of $s$, and different descriptions of the theory may differ by such rescalings.

In section \ref{timed}, we showed that for $\lambda_+>0$, $\lambda_-=0$, the $n$-point function \eqref{absor} vanishes (see the discussion around eq. \eqref{ttone}). Here we note that this feature also follows from \eqref{corr+-}. Indeed, the symmetry under $t\leftrightarrow -t$, $+\leftrightarrow -$ implies that the amplitude \eqref{absor} is proportional to $\lambda_-^{-\frac {i}{p}\sum\omega_j}$. As $\lambda_-$ decreases, this factor leads to a rapidly oscillating phase and the amplitude vanishes in the limit $\lambda_-\to 0$. 

Another calculation from section \ref{timed} that is interesting to generalize to the case $\lambda_\pm>0$ is the comparison of $v_{\rm max}$, \eqref{valvmax}, obtained from the potential \eqref{sppot} and from the amplitudes \eqref{positionspace}. For $\lambda_-=0$ we found a precise agreement between the two. For general $\lambda_-$, the second way of determining $v_{\rm max}$ gives the same answer, \eqref{valvmax}, in terms of the renormalized cosmological constant, since after the redefinition \eqref{renmu} the argument of the function $f$ in \eqref{corr-fin} is the same as before. 

\begin{figure}[t]
\centering
 \includegraphics[width=0.45\textwidth]{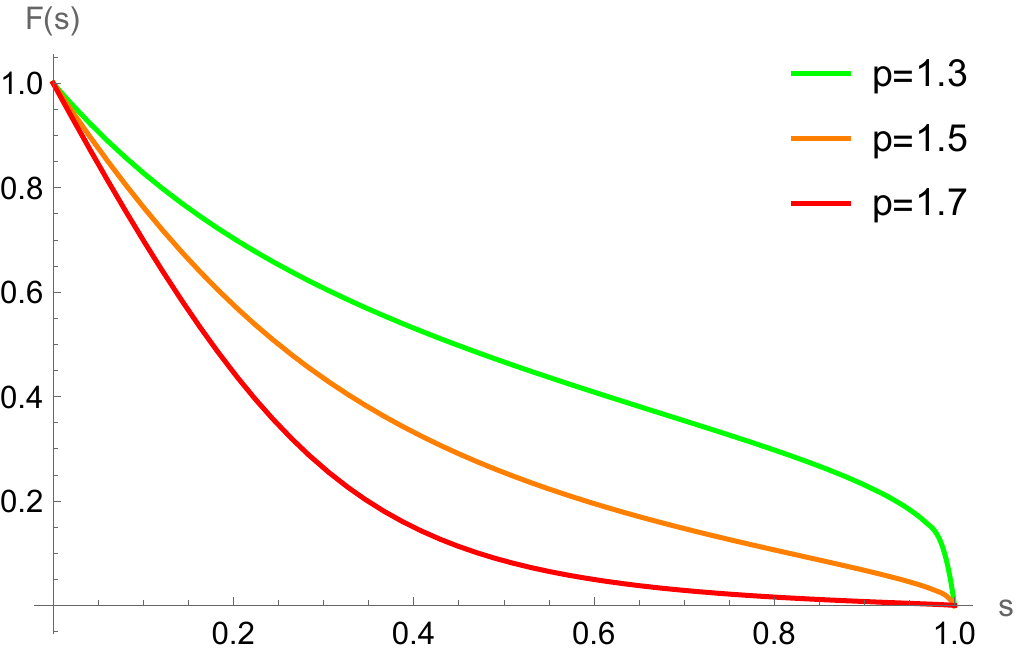}
 \caption{Solutions of \eqref{Fos} for several values of $p$.} 
\label{Fsf1}
\end{figure}

To determine $v_{\rm max}$ from the potential, we need to calculate the value of $v$ for which the Liouville wall associated with \eqref{sppotpm} reaches the speed of light. A short calculation leads to the result 
\ie
\label{defFs}
e^{pv_{\rm max}}=\frac{(p-1)^{p-1}F(s)}{p^p\lambda_+\mu^{p-1}}\ .
\fe
where $F(s)$ is determined by
\ie
\label{Fos}
F(s)\left(1+\frac{p^ps}{(p-1)^p(1-s)^{2-p}F(s)}\right)^{p-1}=1\ .
\fe
For $s=0$, \eqref{Fos} gives $F(0)=1$, and \eqref{defFs} agrees with \eqref{valvmax}, as expected. As $s\to 1$, it behaves as 
\ie
F(s\to 1)\simeq\left(\frac{p}{p-1}\right)^{\frac{p(p-1)}{p-2}}(1-s)^{p-1}~.
\fe
For general $s\in(0,1)$, \eqref{Fos} interpolates between the two behaviors, as demonstrated in figure \ref{Fsf1}. 
Thus, it does not agree with \eqref{valvmax}. The two differ by a redefinition of $\mu$, similar to that in \eqref{renmu},
\ie
\label{newrenmu}
\mu\to\mu G(s)^{\frac{1}{p-1}}~,
\fe
where $G(s)$ is the ratio of the two $F$ functions \eqref{defFs}. This ratio is a smooth finite function for all $s\in [0,1]$. Our view is that agreement up to such a redefinition is sufficient. The difference between the two calculations seems to be due to the fact that one of them (the one from the amplitudes) includes quantum effects in the worldsheet theory, while the other one (based on the form of the Liouville potential) is classical.  

Note that the fact that $v_{\rm max}$ takes the general form \eqref{defFs} follows from the symmetry of the problem under shifts of $t$, $\phi$, with the appropriate rescaling of $\mu$, $\lambda_\pm$, such that the potential \eqref{sppotpm} is invariant. This, together with the freedom to rescale $\mu$ by a function of $s$, \eqref{newrenmu}, seems to make the agreement between the different calculations superfluous. We believe that the main test of the agreement between the different calculations is that, as mentioned above, the resulting ratio $G(s)$ is finite for all $s$. 

\section{Free fermion perspective}
\label{fermis}

The analysis of the previous sections was performed from the perspective of $1+1$ dimensional string theory -- the bulk theory in the holographic correspondence. We did use some results from the dual boundary theory, matrix quantum mechanics (MQM) in a double scaling limit, but only as a tool for computing correlation functions in the standard background \eqref{Sws}. In this section we discuss the physical picture obtained in these earlier sections from the point of view of the boundary theory.  

As reviewed in appendix \ref{ap0}, double scaled MQM can be viewed as a theory of $N\to\infty$ free fermions in an inverted harmonic potential, \eqref{freeH}. The standard background corresponds to a state in which all energy levels up to $-\mu$ are filled (figure \ref{vacuum}), and the Fermi surface takes the form in figure \ref{fes}.  

\begin{figure}[h!]
\centering
 {\subfloat[\label{fseap05}]{\includegraphics[width=0.4\textwidth]{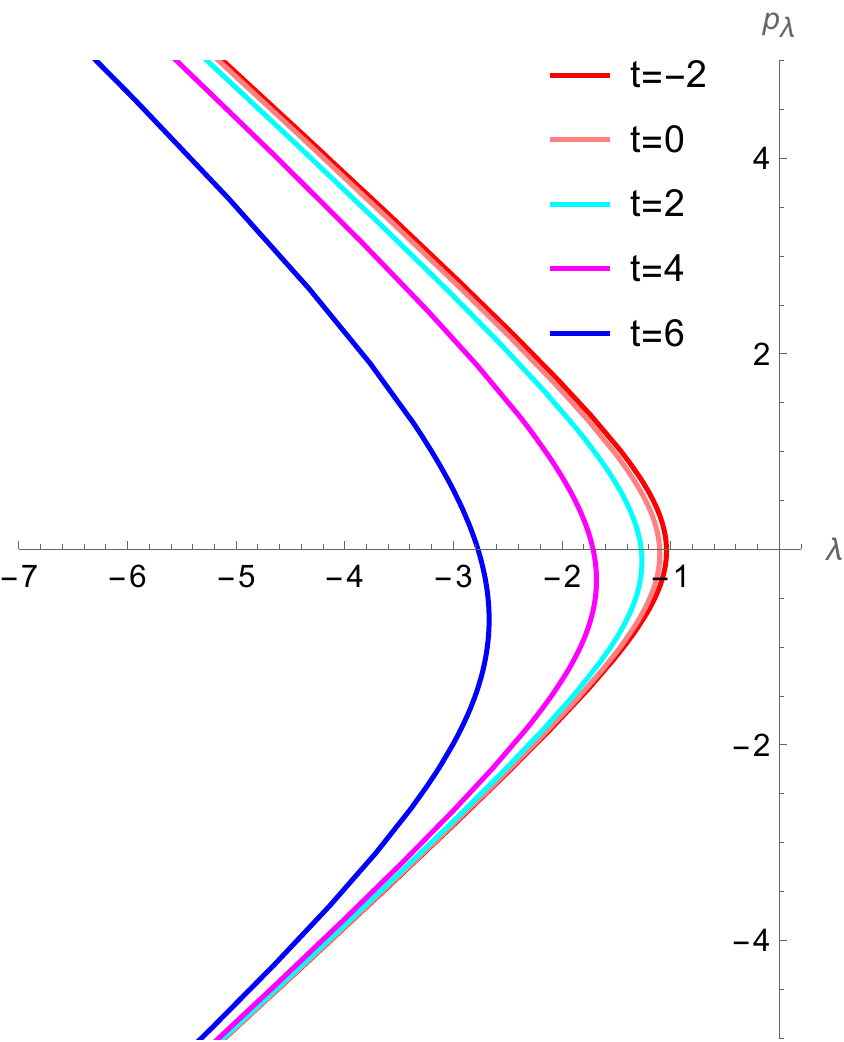}}}~
 \hspace{1.8 cm}
 {\subfloat[\label{fseap15}]{\includegraphics[width=0.4\textwidth]{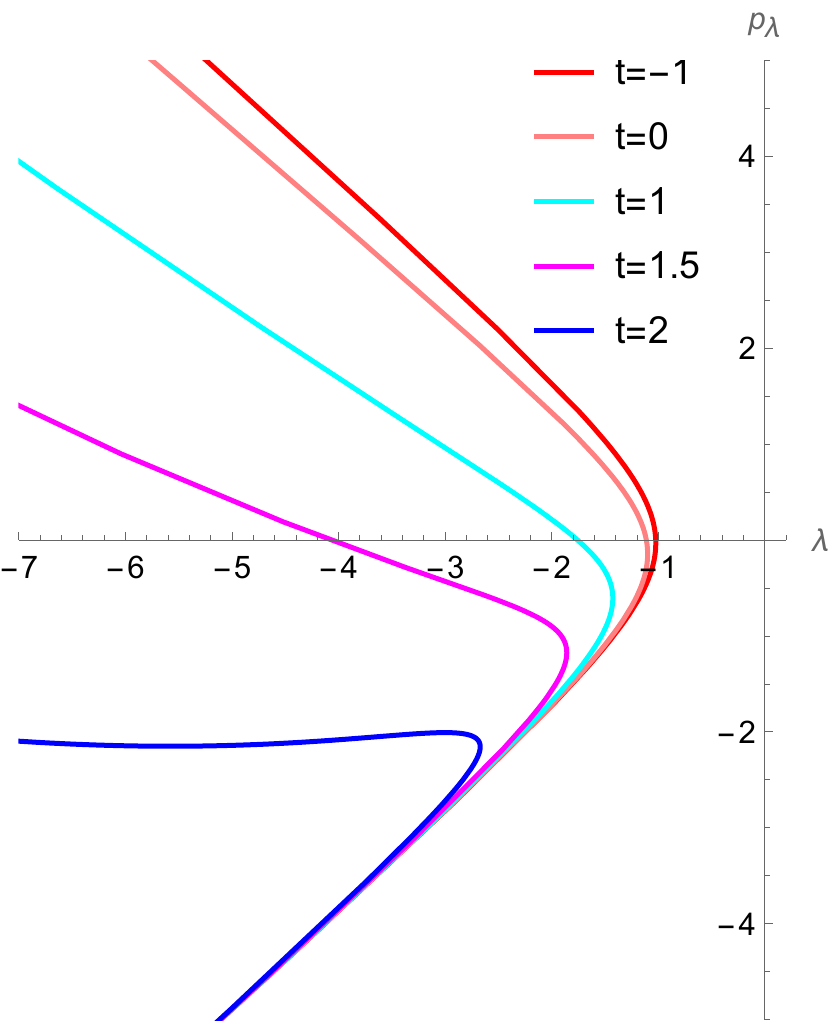}}} 
 \caption{Profile of Fermi sea for (a) $p=0.5$ and (b) $p=1.5$, with $a_0=-1$ and $a_+=-0.1$. In terms of the parametrization \eqref{pfsea}, \eqref{defaone}, $w\to\pm\infty$ correspond to the asymptotic regions of the lower and upper branch, respectively.}
 \label{fsea}
\end{figure}

The backgrounds of this paper correspond to Fermi surfaces that are time-dependent. The dynamics of such Fermi surfaces was studied in 
\cite{Alexandrov:2002fh,Alexandrov:2002pz,Alexandrov:2003uh,Karczmarek:2003pv,Das:2004hw,Das:2004aq,Das:2007vfb}. In particular, the backgrounds \eqref{defL} with general $0<p\le2$ were discussed in 
\cite{Alexandrov:2002fh,Alexandrov:2002pz,Alexandrov:2003uh}. For $\lambda_-=0$, these backgrounds correspond to Fermi surfaces that take the parametric form 
\ie
\label{pfsea}
\lambda=a_0\cosh w+a_+e^{(1-p)w+pt}\ ,\\
p_\lambda=a_0\sinh w+a_+e^{(1-p)w+pt}\ ,
\fe
with $-\infty<w<\infty$, and 
\ie
\label{defaone}
a_0= -\sqrt{2\mu}~, \qquad a_+= -\frac{1}{\sqrt{2}}\lambda_+\mu^{\frac{p-1}{2}}~.
\fe
Eliminating $w$ in \eqref{pfsea}, the Fermi surface takes the form\footnote{Note that \eqref{pfsea}, \eqref{defaone} imply that $p_\lambda>\lambda$ for all $p$, $w$.}
\ie
\label{pfsea1}
\lambda^2-p_\lambda^2+\frac{2a_+}{(-a_0)^{p-1}}e^{pt}(p_\lambda-\lambda)^p=a_0^2~.
\fe
In figure \ref{fsea}, we plot \eqref{pfsea1} for $p=0.5$ and $p=1.5$, to illustrate the time evolution of the Fermi surface for $p<1$ and $p>1$, respectively. At early times, $t\to -\infty$, the Fermi surface is approximately static, given by the $\lambda<0$ branch of the hyperbola $\lambda^2-p_\lambda^2=2\mu$. The fermions fill the region  $-\infty<\lambda\le-\sqrt{2\mu}$ in $\lambda$ space. In terms of the coordinate $\phi$, related to $\lambda$ via the relation~\cite{Polchinski:1994mb}
\ie
\label{lamphi}
\lambda=-\sqrt{2}e^{-\phi}~,
\fe
it is $-\infty<\phi\le -\frac{1}{2}\ln \mu$. 

As $t$ increases, the Fermi surface  moves to the left. Its rightmost edge, $\lambda_{\rm max}(t)$, can be determined by solving the equation $\partial_w\lambda=0$. At large $t$ and $p<1$ one finds, from \eqref{pfsea}, 
$w\sim -\frac{p}{2-p}t$, $\lambda_{\rm max}\sim e^{\frac{p}{2-p}t}$, and \eqref{lamphi}, $\phi_{\rm max}\sim -\frac{p}{2-p}t$. So, the edge of the distribution  moves to the left with speed $\frac{p}{2-p}$, which is smaller than the speed of light. Note that this velocity is the same as that of the Liouville wall (see figure \ref{2powa}), though we will see that the two (the edge of the Fermi surface and the Liouville wall) are distinct objects. For $p>1$, $\lambda_{\rm max}(t)$ corresponds at large $t$ to $w\sim t$. So $\lambda_{\rm max}\sim e^t$, and $\phi_{\rm max}\sim -t$. Thus, in this case the edge of the Fermi surface moves to the left with the speed of light. 

The lower branch of the Fermi surface in figure \ref{fsea} corresponds asymptotically to $w\to\infty$ in \eqref{pfsea}. In this limit, $p_\lambda-\lambda=|a_0|e^{-w}\to 0$, so the Fermi surface approaches the line $p_\lambda=\lambda$ for all $t$. The asymptotic region of the upper branch in figure \ref{fsea} corresponds to $w\to-\infty$. Using the fact that $p_\lambda+\lambda=a_0e^w+2a_+e^{(1-p)w+pt}$, we see that for $p<1$, $p_\lambda+\lambda\to 0$ in this limit, while for $p>1$,  $p_\lambda+\lambda\to -\infty$.\footnote{Note, however, that the sum $p_\lambda+\lambda$ grows slower with $|w|$ than each of the two quantities separately.} 

The massless tachyon in the bulk $1+1$ dimensional string theory description corresponds in the fermion language to a ripple on the Fermi surface. In the standard background, which is described by the Fermi surface of figure \ref{fes}, an incoming tachyon is described by a ripple that starts at early time on the upper branch of the hyperbola, at large negative $\lambda$ (or $\phi$, \eqref{lamphi}). Since $p_\lambda$ is positive there, the ripple propagates to the right. It corresponds to $\mathcal{T}^+$ in figure \ref{1powa}. 

As time goes by, the ripple propagates down the Fermi surface of figure \ref{fes}.  Eventually, at some time, it reaches the $\lambda$ axis, after which it moves to the lower branch of the hyperbola. There, it has $p_\lambda<0$, and thus is moving to the left. It corresponds (asymptotically, at large $t$) to $\mathcal{T}^-$. The transition between the two regimes happens at $p_\lambda=0$, \IE $\lambda=-\sqrt{2\mu}$, and \eqref{lamphi} $\phi=-\frac12\ln\mu$. The latter is precisely the location of the Liouville wall (see figure \ref{1powa}), as expected.

Clearly, the above description can be generalized to the case of non-zero $\lambda_+$. In this case, the incoming tachyons correspond to ripples starting at early times in the upper left region in figure \ref{fsea}, propagate to the right, and change direction at the value of $\lambda$ where the time-dependent Fermi surface intersects the $\lambda$ axis. Thus, the position of the Liouville wall is given by setting $p_\lambda=0$ in the expression for the Fermi surface \eqref{pfsea1}. Using \eqref{sppot}, \eqref{defaone}, \eqref{lamphi}, one can check that this gives precisely the equation $V_{\rm st}(t,\phi)=1$, with $V_{\rm st}$ given by \eqref{sppot}. 

In the previous sections, we saw that there is a qualitative difference between the case $p<1$, where the trajectory of the Liouville wall remains timelike for all $t$, and $p>1$, where it eventually becomes spacelike. It is interesting to see how this difference manifests itself in the free fermion language. 

For this purpose, we turn to a more detailed description of the ripples on the Fermi surface, that correspond in this language to tachyon perturbations in the $1+1$ dimensional string theory. Small ripples can be thought of as points on the Fermi surface that follow the trajectories~\cite{Polchinski:1991uq} 
\ie
p_\lambda=\dot{\lambda}~,\qquad \lambda=\dot{p}_\lambda\ .
\fe
The solution of these equations is 
\ie
\label{dyn}
\lambda=-\rho \cosh (t-\sigma)\ ,\qquad p_\lambda =-\rho\sinh (t-\sigma)\ ,
\fe
where $\rho$ and $\sigma$ are determined by the initial conditions. We will take $\rho>0$, since we want to consider perturbations on the Fermi surface that extends to $\lambda\to-\infty$. 

\begin{figure}[h]
\centering
 \includegraphics[width=0.45\textwidth]{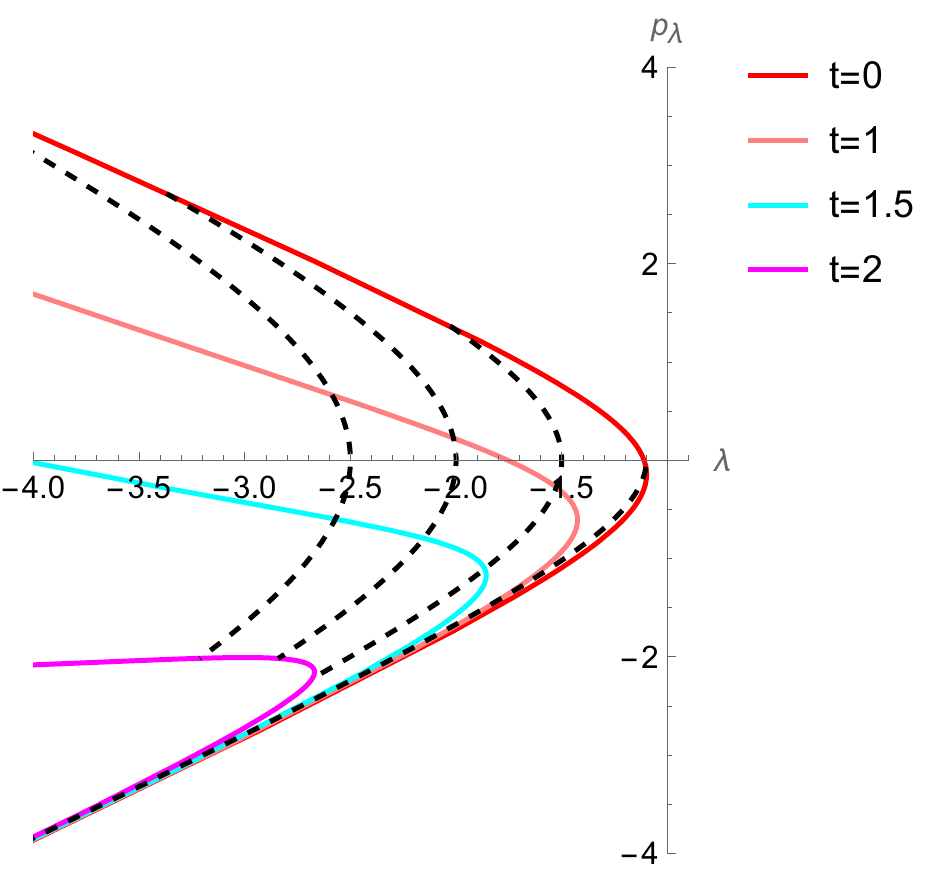}
 \caption{The solid lines describe the form of the Fermi surface for $p=1.5$, $a_0=-1$, $a_+=-0.1$ and a few values of $t$. The dashed black lines describe ripples with various values of $\rho$.} 
\label{ripp}
\end{figure}

The parameter $\sigma$ is not independent of $\rho$, since the trajectories in question must lie on the Fermi surface. Plugging \eqref{dyn} in \eqref{pfsea1}, we get
\ie
\label{tturn}
\sigma(\rho)=\frac{1}{p}\ln \frac{(\rho^2-a_0^2)(-a_0)^p}{2a_0a_+\rho^p}\ .
\fe
The trajectories \eqref{dyn}, \eqref{tturn} have the following qualitative structure. As $t\to-\infty$, eq. \eqref{dyn} implies that $\lambda\to-\infty, ~p_\lambda\to\infty$, with $p_\lambda\sim -\lambda$. Thus, at early times \eqref{dyn} describes a ripple moving to the right from large negative $\lambda$, with speed close to the speed of light. At $t=\sigma(\rho)$, \eqref{tturn}, the ripple turns around, and starts going to the left (\IE $p_\lambda$ changes sign). At that time, the ripple is at $\lambda=-\rho$, which is thus the position of the Liouville wall at $t=\sigma(\rho)$. As $t\to\infty$, $\lambda,~ p_\lambda\to -\infty$, with $p_\lambda\sim \lambda$, \IE the speed of the ripple approaches the speed of light again. We plot a few examples of these trajectories in figure \ref{ripp}. 

Eq. \eqref{tturn} implies that the constant $\rho$ varies from $|a_0|$ to infinity. As $\rho\to|a_0|$, $\sigma\to-\infty$, and \eqref{dyn} describes a trajectory which coincides with the shape of the early-time Fermi surface. The trajectory crosses the $\lambda$ axis at $t=\sigma$, which also goes to $-\infty$ in this limit. Thus, this limit describes ripples that are reflected from the time-dependent Liouville wall at a very early time. 

For such ripples, at finite times, \EG $t=0$, $\lambda=-\rho \cosh(-\sigma)\sim a_0e^{-\sigma}\to-\infty$, which is much larger (in absolute value) than $\lambda_{\rm max}$ at that time. Furthermore,  $p_\lambda\sim \lambda$ at that time, which means that the ripple approaches the speed of light. This corresponds to a ripple that propagates to the left along the lower branch of the Fermi surface in figure \ref{fsea}, and is well separated from the tip of the Fermi surface. 

For large $\rho$, \eqref{tturn} implies that $\sigma\sim \frac{2-p}{p}\ln\rho$. Such ripples are reflected from the Liouville wall at a late time, $t=\sigma$. The fate of these ripples depends on the sign of $p-1$. To see that, it is useful to ask at what time does a ripple with given $\rho$ pass the tip of the Fermi surface. This time is obtained by solving the equation
\ie
\label{plpr}
\partial_\rho\lambda|_{t=t_0}=-\cosh (t_0-\sigma(\rho))+\rho \sinh (t_0-\sigma(\rho))\sigma'(\rho)=0\ ,
\fe
which gives
\ie
\label{tzero}
e^{2t_0}=\frac{1}{(1-p)\rho^2+a_0^2p}\left(\frac{(\rho^2-a_0^2)(-a_0)^p}{2a_0a_+}\right)^{\frac{2}{p}}\ .
\fe
As a check of this equation, it is clear from figures \ref{fsea}, \ref{ripp}, that it must be that $t_0(\rho)>\sigma(\rho)$, \IE a ripple with given $\rho$ first encounters the Liouville wall, turns around, and at a later time approaches the tip of the Fermi surface. One can check that for all $\rho$ for which \eqref{tzero} has a real solution, this is indeed the case.

As $\rho\to |a_0|$, $t_0(\rho)\to-\infty$. This is consistent with what we discussed above, where we saw that for $\rho\to |a_0|$, the ripple crosses the $\lambda$ axis at $t=\sigma\to -\infty$, \eqref{tturn}. At that early time, the hyperbola describing the Fermi surface is symmetric under $p_\lambda\to-p_\lambda$ (the Fermi surface is approximately static), so the time at which $p_\lambda=0$ approximately coincides with the tip of the Fermi surface, $\sigma(\rho)\sim t_0(\rho)$. 

As $\rho$ increases, $t_0(\rho)$, \eqref{tzero}, increases monotonically. For $p<1$, as $\rho$ varies from $|a_0|$ to $\infty$, $t_0$ runs from $-\infty$ to $+\infty$. In other words, a ripple with any $\rho$ starts on the upper branch of the Fermi surface at early time, 
reaches the tip of the Fermi surface at some finite time $t_0(\rho)$, \eqref{tzero}, and then moves to the lower branch of the Fermi surface in figure \ref{fseap05}. The S-matrix discussed in section \ref{timed} for this case corresponds to such processes. 

On the other hand, for $p>1$, as we increase $\rho$, we get to a critical $\rho$, 
\ie
\label{rhoc}
\rho_c= \sqrt{\frac{p}{p-1}}|a_0|~,
\fe
at which $t_0$ \eqref{tzero} diverges. For $|a_0|<\rho<\rho_c$, the picture is like for $p<1$, as illustrated by the two rightmost dashed lines in figure \ref{ripp}. But for $\rho>\rho_c$, the ripple never gets to the tip of the Fermi surface, and stays on the upper branch for all $t$. This is the case for the trajectories described by the two leftmost dashed lines in figure \ref{ripp}. For $\rho=\rho_c$, the ripple approaches the tip of the Fermi surface as $t\to\infty$.

Interestingly, the bound $\rho<\rho_c$ described above corresponds to the bound on $v$, \eqref{valvmax}, encountered in section \ref{timed} above. Indeed, we can calculate the value of $v=t+\phi$ at which a ripple with $\rho=\rho_c$ is reflected from the Liouville wall. The time at which it is reflected is $t_c=\sigma(\rho_c)$ \eqref{tturn}. The spatial position is $\lambda_c=-\rho_c$ (see the discussion after \eqref{dyn}), or $\phi_c=-\ln \frac{\rho_c}{\sqrt{2}}$, \eqref{lamphi}. One can check that $t_c+\phi_c=v_{\rm max}$ given in \eqref{valvmax}. 

Ripples with $\rho<\rho_c$ are reflected at $v(\rho)<v_{\rm max}$. As expected from the above discussion, the function $v(\rho)$, has a maximum at $\rho=\rho_c$. This can be seen as follows. We have 
\ie
\label{vofrho}
v(\rho)=t(\rho)+\phi(\rho)=\sigma(\rho)-\ln\frac{\rho}{\sqrt2}~.
\fe
Thus, $v'=0$ is equivalent to 
\ie
\label{maxim}
\sigma'(\rho)=\frac{1}{\rho}~.
\fe
To see that this relation is satisfied at $\rho=\rho_c$, we look back at \eqref{plpr}, recalling that as $\rho\to\rho_c$, $t_0\to\infty$. Conversely, if \eqref{maxim} is satisfied at some $\rho$, and $\sigma(\rho)$ is finite, $t_0$ must go to infinity, which means that $\rho=\rho_c$, \eqref{tzero}.  

Therefore, $v(\rho)$ has the following qualitative structure. It monotonically increases with $\rho$ until the critical value \eqref{rhoc}, at which it is given by $v_{\rm max}$, \eqref{valvmax}, and then monotonically decreases for $\rho>\rho_c$. This is precisely the behavior one expects from the form of the Liouville wall of figure \ref{2powa2}.

For $\rho<\rho_c$, \IE $v<v_{\rm max}$, we have a picture similar to that for $p<1$. In particular, it appears that for ripples with these values of $\rho$ there is an S-matrix, describing the propagation from asymptotic past infinity on the upper branch to asymptotic future infinity on the lower one. We will return to these observables in the next section. Here, we want to mention two things about them. One is that they are very similar to the position space observables we defined in section \ref{timed}, \EG in eq. \eqref{positionspace}.\footnote{The direct analog of the discussion here is the two point function $\langle \mathcal{T}^+\mathcal{T}^-\rangle.$} In particular, the range of $v$ and the physical interpretation are very similar in the two cases. 

The second is that in both cases there is a puzzle associated with these observables. In the language of this section, an incoming ripple that is reflected from the Liouville wall at a $v(\rho)<v_{\rm max}$, can propagate to the left on the lower branch of the Fermi surface. However, eventually it is  overtaken by the Liouville wall, which corresponds to the intersection of the Fermi surface with the $\lambda$ axis in figure \ref{fseap15}, and asymptotically moves faster than light. Thus, their physical interpretation as string theory observables is unclear, since the latter should be defined far from the Liouville wall. 

\begin{figure}[h!]
\centering
 {\subfloat[\label{fsea2p05}]{\includegraphics[width=0.4\textwidth]{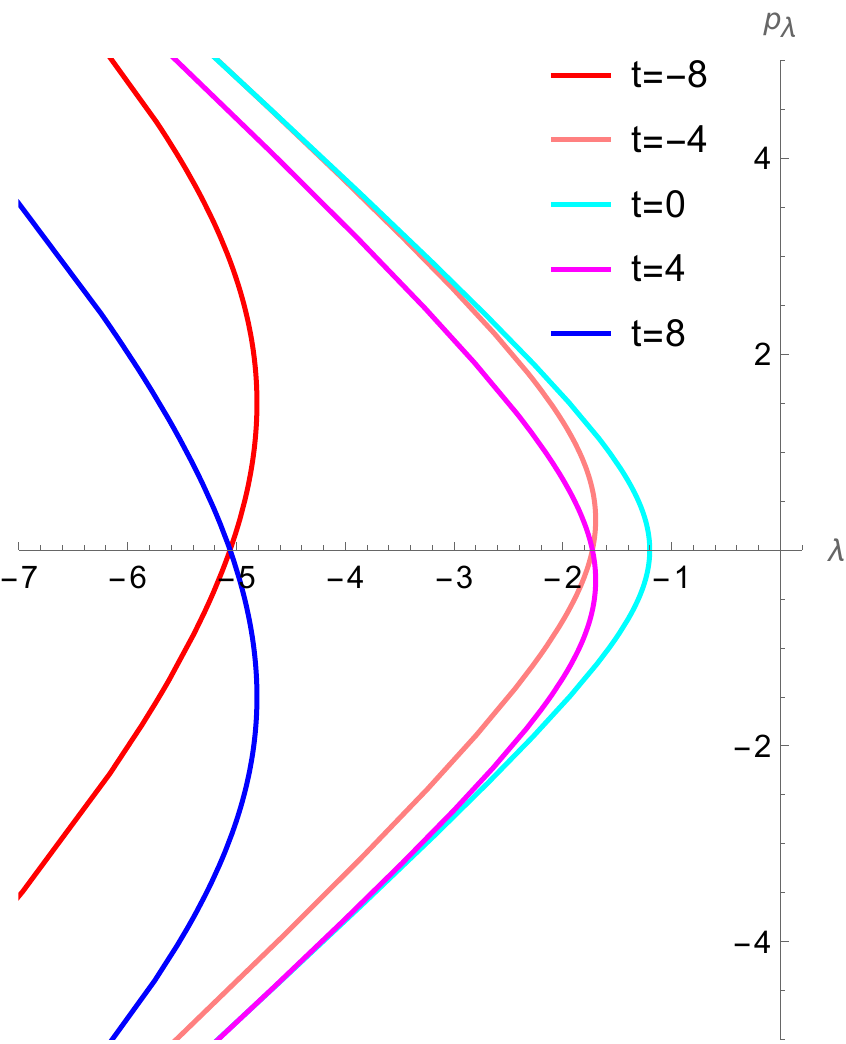}}}~
 \hspace{1.8 cm}
 {\subfloat[\label{fsea2p15}]{\includegraphics[width=0.4\textwidth]{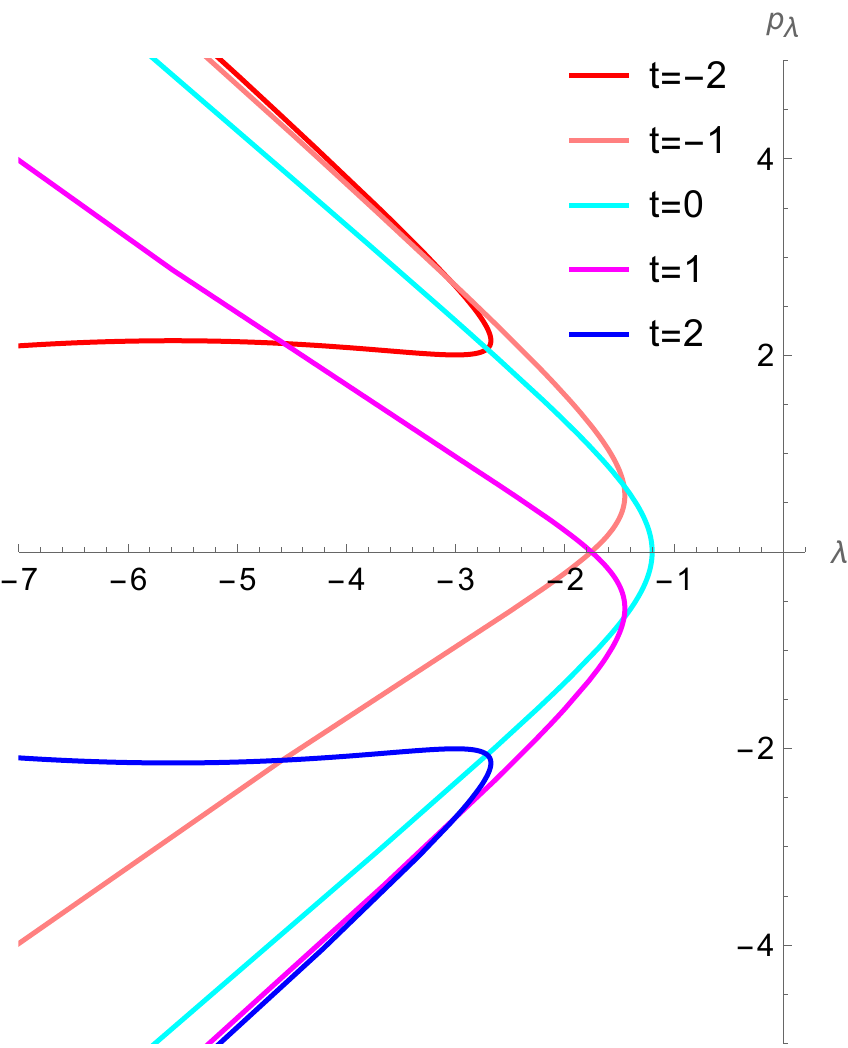}}} 
 \caption{The profile of the Fermi sea \eqref{pfsea2} for (a) $p=0.5$ and (b) $p=1.5$, with $a_1=-1$ and $a_\pm=-0.1$. Note the symmetry under $t\to -t$.}
 \label{fsea2}
\end{figure}

In section \ref{genz} we discussed the generalization of the above analysis to the case where both $\lambda_+$ and $\lambda_-$ are non-zero. It is interesting to describe this case in the free fermion language. The Fermi surface takes now the form \cite{Alexandrov:2003uh}
\ie
\label{pfsea2}
\lambda=a_1\cosh w+a_+e^{(1-p)w+pt}+a_-e^{-(1-p)w-pt}\ ,\\
p_\lambda=a_1\sinh w+a_+e^{(1-p)w+pt}-a_-e^{-(1-p)w-pt}\ ,
\fe
where $a_\pm=-2^{-\frac{p}{2}}\lambda_\pm |a_1|^{p-1}$ and $a_1$ satisfies the equation
\ie
\label{a1}
\mu=\frac{a_1^2}{2}+(1-p)\lambda_+\lambda_-\left(\frac{a_1^2}{2}\right)^{p-1}\ .
\fe
Eqs. \eqref{pfsea2}, \eqref{a1} generalize 
\eqref{pfsea}, \eqref{defaone} to the case where both $\lambda_+$ and $\lambda_-$ are non-zero. In figure \ref{fsea2} we plot the resulting Fermi surfaces.

To understand \eqref{a1}, it is useful to define a quantity $\psi$, via the equation
\ie
\label{defpsi}
\frac{a_1^2}{2}=\mu\psi\ .
\fe
Plugging \eqref{defpsi} into \eqref{a1}, and comparing to \eqref{mapsprime}, we see that 
\ie
\label{formpsi}
\psi=\frac{1}{1-s}~.
\fe
Thus, one can think of $\frac{a_1^2}{2}$ as the renormalized cosmological constant discussed in section \ref{genz}, around eq. \eqref{renmu}.  

The qualitative picture is expected to be the same for this case, as in the discussion of the special case $\lambda_-=0$ above. For $p<1$, ripples on the Fermi surface propagate from the upper branch of the hyperbolas in figure \ref{fsea2p05} at early times to the lower branch at late times, and one can define an S-matrix. For $p>1$, there is a maximal value of $v$, $v_{\rm max}$, beyond which they remain trapped, and do not make it to asymptotic infinity on the lower branch. We next calculate $v_{\rm max}$.

Recall that $v(\rho)$ is the value of $t+\phi$ at the point where the ripple with a certain $\rho$ is reflected from the Liouville wall. As $\rho$ approaches its critical value, $v$ approaches $v_{\rm max}$, and the time at which the ripple passes the edge of the Fermi surface goes to infinity. This can be used to calculate $v_{\rm max}$ for the general case \eqref{pfsea2}. 

The tip of the Fermi surface can be obtained by solving the equation 
\ie
\label{soltip}
\frac{\partial \lambda}{\partial w}=a_1\sinh w+(1-p)a_+e^{(1-p)w+pt}-(1-p)a_-e^{-(1-p)w-pt}=0\ 
\fe
for $w$, and plugging back into \eqref{pfsea2}. To calculate $v_{\rm max}$, we need to take $t\to\infty$. In this limit, the solution of \eqref{soltip} is
\ie
\label{solwww}
w\simeq t+\frac{1}{p}\ln \frac{2(p-1)a_+}{a_1}~.
\fe
Plugging \eqref{solwww} into  \eqref{pfsea2}, we get
\ie
\lambda(t)\simeq-\frac{e^t}{2}\left(\frac{2p^p|a_1|^{p-1}|a_+|}{(p-1)^{p-1}}\right)^{\frac{1}{p}}\ .
\fe
On the other hand, from \eqref{dyn} we learn that at large $t$,
\ie
\lambda(t)\simeq-\frac{e^t}{2}\rho_ce^{-\sigma_c}\ .
\fe
Comparing the two equations, we have
\ie
\label{rhosig}
\rho_ce^{-\sigma_c}=\left(\frac{2p^p|a_1|^{p-1}|a_+|}{(p-1)^{p-1}}\right)^{\frac{1}{p}}\ .
\fe
The expression on the l.h.s. of \eqref{rhosig} has a nice interpretation. As discussed earlier in this section, the critical ripple is reflected from the Liouville wall at $(t,\lambda)=(\sigma_c, -\rho_c)$, or in terms of $(t,\phi)$ at $(t_c,\phi_c)=(\sigma_c, -\ln\frac{\rho_c}{\sqrt2})$.  Thus,
\ie
\label{rcsc}
e^{v_{\rm max}}=e^{\phi_c+\sigma_c}=\frac{\sqrt{2}}{\rho_c}e^{\sigma_c}\ .
\fe
Plugging \eqref{rcsc} in \eqref{rhosig}, we conclude that
\ie
e^{v_{\rm max}}=\sqrt{2}\left(\frac{(p-1)^{p-1}}{2p^p|a_1|^{p-1}|a_+|}\right)^{\frac{1}{p}}\ ,
\fe
or
\ie
\label{epvfermi}
e^{pv_{\rm max}}=\frac{(2p-2)^{p-1}}{p^p\lambda_+a_1^{2p-2}}\ .
\fe
This is the same result we got from an analysis of the amplitudes in section \ref{genz}, and as there, it also agrees with the result we got for $\lambda_-=0$, \eqref{valvmax}, written in terms of the renormalized cosmological constant \eqref{defpsi}.

\section{Summary and discussion}
\label{discuss}

\subsection{Summary}

The main goal of this paper was to study $1+1$ dimensional string theory in time-dependent backgrounds, building on previous work on this subject. From the bulk point of view in the holographic duality between $1+1$ dimensional string theory and double scaled matrix quantum mechanics, these backgrounds correspond to solutions in which the Liouville wall is accelerating from one velocity in the far past to another in the far future. In the boundary theory, they correspond to solutions in which the Fermi surface of free fermions in an inverted harmonic potential is time-dependent. 

We discussed different types of backgrounds of this kind. In the background we studied in detail in section \ref{timed},  the Liouville wall moves towards the boundary, with a
velocity that goes from zero as $t\to-\infty$,  to a finite value as $t\to\infty$. The final velocity of the wall can be smaller or larger than the speed of light. In the former case, depicted in figure \ref{2powa1}, one can define an S-matrix for $n$ incoming tachyons to go to $n'$ tachyons. We used a Wick rotation from Euclidean spacetime to study some particular processes: $n\to 1$ scattering \eqref{nto1lambge0}, and production of $n$ outgoing tachyons \eqref{finminus} in the time-dependent background.  

When the final velocity of the Liouville wall is larger than the speed of light, asymptotic future null infinity is shielded by the Liouville wall, and the outgoing massless tachyons can no longer be defined. Nevertheless, we found that one seems to be able to define asymptotic observables, \eqref{positionspace} with $v_j<v_{\rm max}$, the latter given by \eqref{valvmax}. One can think of these observables as associated with outgoing tachyons produced in the time-dependent background. The bound on the null coordinate  $v=t+\phi$ is related to the fact that these tachyons are produced at the Liouville wall, which in this case has a maximal value of $v$, see figure \ref{2powa2}. 

In section \ref{genz} we discussed a generalization of the above system, in which the velocity of the Liouville wall approaches (equal and opposite) finite values at early and late times (see figure \ref{3powa}). When these velocities are smaller than the speed of light, we found a similar structure to that of section \ref{timed}. One can again define an $n\to n'$ S-matrix for tachyon scattering, and compute it using matrix model and worldsheet techniques. We demonstrated this by computing the amplitude for creation of $n$ tachyons, given in eq. \eqref{corr+-}. These amplitudes are qualitatively similar to those encountered in section \ref{timed}, \eqref{finminus}, and become even more so after a rescaling of $\mu$ given in \eqref{renmu}. 

A new feature of these amplitudes is a divergence at a finite value of the coupling \eqref{newmaps}, that determines the local acceleration of the Liouville wall. In the closely related Euclidean problem, the corresponding  divergence signals the decoupling of the Euclidean time $X$ from the worldsheet dynamics -- a kind of dynamical dimensional reduction. We proposed that similarly, the  divergence of \eqref{corr+-} may be due to a decoupling of time, though a more complete understanding is required. 

In section \ref{fermis} we discussed the time-dependent backgrounds described above from the point of view of the dual matrix quantum mechanics. In the double scaling limit this theory reduces to a theory of free fermions, and different backgrounds correspond to different choices of the shape of the Fermi surface. Tachyon perturbations in the bulk $1+1$ dimensional string theory are described in this language by ripples on the Fermi surface. 

We presented the Fermi surfaces corresponding to the backgrounds of sections \ref{timed}, \ref{genz}, and studied the dynamics of ripples on these surfaces as a function of the parameters. We found a nice agreement between the picture based on scattering amplitudes in sections \ref{timed}, \ref{genz}, and the free fermion description. 

In the region of parameter space where the Liouville wall follows a timelike trajectory, we found that ripples on the Fermi surface propagate from the upper branch in figure \ref{fsea2p05} at early times to the lower branch at late times, thus giving an analog of the $1\to 1$ S-matrix of section \ref{timed}. We showed that their trajectories can be viewed as due to reflection from a time-dependent wall, that coincides with the Liouville wall discussed in section \ref{intro}. We also showed that the Fermi surface picture only exists below some critical value of the coupling \eqref{newmaps}, which agrees precisely with that found in section \ref{genz}. 

In the region where the Liouville wall turns spacelike at early and/or late times, we showed that the trajectories of ripples on the Fermi surface make it to the corresponding asymptotic infinity only for some values of the parameters describing the ripple. We found that the bounds on these parameters are the same as those found from a seemingly very different point of view, by studying the amplitudes in sections \ref{timed}, \ref{genz}.

In summary,  we found a nice agreement between three, seemingly quite different, points of view on tachyon dynamics in the time-dependent backgrounds we studied: (1) scattering from the time-dependent Liouville wall \eqref{sppot}, \eqref{sppotpm}; (2) amplitudes obtained by Wick rotation from the Euclidean correlation functions; (3) the dynamics of ripples on the time-dependent Fermi surfaces of section \ref{fermis}. 

Much remains to be done. We next list some issues that require further attention.

\subsection{Observables for $p>1$}

One of the surprising results of our analysis was that in backgrounds where a priori there should not be asymptotic observables associated with past and/or future asymptotic infinity, we nevertheless found such observables. Moreover, we found them from two different points of view: the study of scattering amplitudes in sections \ref{timed}, \ref{genz}, and the study of ripples on a time-dependent Fermi surface in section \ref{fermis}. 

As an example, in the analysis of section \ref{pgone}, we found that one can define observables, \eqref{positionspace}, which depend on lightlike position variables $v_j$, that are bounded from above by $v_{\rm max}$, \eqref{valvmax}. One can think of these observables as describing the emission of tachyons from the time-dependent Liouville wall, and of $v_{\rm max}$ as the largest value of $v=t+\phi$ along the wall (figure \ref{2powa2}). In the analysis of section \ref{fermis}, these observables correspond to ripples \eqref{dyn}, with $\rho<\rho_c$ \eqref{rhoc}, which start at early times on the upper branch of the hyperbola of figure \ref{fseap15} and make it at late times to the lower branch, along which they propagate to asymptotic infinity. 

The existence of these observables is at first sight puzzling. As mentioned above, in terms of figure \ref{2powa2} they appear to correspond to outgoing tachyons, $\mathcal{T}^-_\omega$, emitted by the accelerating Liouville wall. However, regardless of the value of $v$ at which they were emitted, eventually the Liouville wall catches up with them, and they are absorbed by it. In terms of the picture of section \ref{fermis}, they correspond to ripples propagating to the left on the lower branch of the hyperbolas in figure \ref{fseap15}. In that picture, the Liouville wall corresponds to the intersection of the hyperbolas with the $\lambda$ axis, which asymptotically, at late times, moves faster than light, and thus eventually these ripples find themselves behind the wall. 

This raises the question how we can define the $\mathcal{T}^-$ observables, when they do not make it to the future region well outside  the Liouville wall. We will leave a detailed analysis of this issue to future work, but a possible resolution of this tension is the following. It seems clear that the operators $\mathcal{T}^-_\omega$ defined in \eqref{voL} {\it with real $\omega$} do not make sense when future null infinity is behind the Liouville wall. However, it is possible that these observables can be defined when $\omega$ has a finite imaginary part. This imaginary part must be taken to be positive, since otherwise the corresponding operators \eqref{voL} are normalizable, which would imply the existence of normalizable states, that are in general not there. 

If we write $\omega=w_r+i\omega_i$, with $\omega_i>0$, the resulting operators $\mathcal{T}^-_\omega$, behave at late times like $\exp(-\omega_it)$, \IE they go exponentially to zero at late times, which is consistent with the picture suggested by figure \ref{2powa2}, that they correspond to tachyons that penetrate the Liouville wall at late times. It also resolves the puzzle regarding the definition of these observables, since they are now defined at the boundary of the spacetime, $\phi\to-\infty$, which remains outside the Liouville wall for all $t$. 

The proposal that the late time observables decay exponentially as $t\to\infty$ suggests that the system with an asymptotically spacelike Liouville wall has a unique final state. This is reminiscent of the proposal by \cite{Horowitz:2003he} of a unique final state associated with the black hole singularity, and of the Hartle-Hawking proposal for a unique wavefunction of the universe, which is associated to the Big Bang singluarity \cite{Hartle:1983ai}. 

This proposal is also reminiscent of \cite{Elitzur:2002rt}, who studied string dynamics in backgrounds with cosmological singularities. These backgrounds contain big bang and big crunch singularities, and thus one cannot define standard string theory S-matrix observables.  At the same time, they are described by solvable worldsheet theories, which contain physical observables that can be studied using standard worldsheet techniques. 

The authors of \cite{Elitzur:2002rt} showed that the resolution of this tension is that the cosmological spacetimes they studied contain additional regions, referred to as  ``whiskers'', and the natural string observables are non-normalizable vertex operators defined at the boundaries of these regions, where they approach linear dilaton spacetimes. 

The setup of \cite{Elitzur:2002rt} is similar to the one depicted in our figure \ref{3powa} with $p>1$. Like there, the Liouville wall of figure \ref{3powa} shields the past and future null infinities, and thus provides soft versions of big bang and big crunch singularities. Therefore, one cannot define a standard S-matrix but, according to our proposal above, one can define non-normalizable observables at the timelike boundary $\phi\to-\infty$, which is the analog of the boundary of one of the whiskers for this case. 

One advantage of the systems described in this paper over those in \cite{Elitzur:2002rt} is that in the ones here one can study the behavior both when there is no spacelike singularity $(p<1)$ and when there is one $(p>1)$. In contrast, in the system of \cite{Elitzur:2002rt}, a past and future spacelike singularity is always present.

If the proposal for the observables we made before for $p>1$ is correct, one needs to understand its implications for the Fermi surface picture of section \ref{fermis}. In particular, one needs to understand how to see the fact that the observables correspond to non-normalizable operators that decay exponentially at early and/or late time. It may be that to do that one will have to generalize the description of ripples on the Fermi surface as points that follow the trajectories \eqref{dyn} to a study of the dynamics of finite size ripples. We will leave all these issues to future work.

Another puzzle that may require going beyond the description of tachyons in $1+1$ dimensional string theory as pointlike ripples on the Fermi surface is the following. In section \ref{genz} we showed that when the coupling \eqref{newmaps} increases, we encounter a singularity of the amplitudes \eqref{corr+-}, \eqref{corr-fin} at a finite value of the coupling. In section \ref{fermis} we argued that in the fermion language this corresponds to the fact that the time-dependent Fermi surface \eqref{pfsea2}, \eqref{a1} only exists in some region in parameter space, and showed that this region is the same as that in section \ref{genz}. However, if one looks at the critical Fermi surface, by plugging $s=1/(p-1)$ in \eqref{formpsi}, one finds a smooth Fermi surface, that does not show any signs of the divergences encountered in section \ref{genz}. It is possible that to understand these divergence in the language of section \ref{fermis}, one needs to study the dynamics of finite size ripples. Another possibility, suggested by the form of the divergences in eq. \eqref{corr+-}, is that they  are a feature of the vacuum, and are insensitive to the perturbations. This too will be left to future work.

\subsection{Properties of scattering amplitudes}

The string scattering amplitudes computed in this work relied on a worldsheet analysis of the background \eqref{defL}. Indeed, the MQM was used only as a computational tool to evaluate string scattering amplitudes in the undeformed $1+1$ dimensional string theory. In section \ref{fermis}, we reviewed the Fermi surface picture that is dual to the time-dependent background \eqref{defL}, but used it only to study properties of the time-dependent Liouville wall, and the trajectories of pointlike ripples. It would
be interesting to use this free fermion description to compute the scattering amplitudes \eqref{nto1int}, \eqref{finminus} for $p<1$. This would be a non-trivial check of the procedure to compute the scattering amplitudes followed in this work, as well as of the proposed Fermi surface dual to the time-dependent background. We leave this question to future work.

In the region of parameter space for which $p<1$, the scattering amplitudes should satisfy constraints coming from unitarity and causality. In time-dependent backgrounds, these constraints are not straightforwardly implemented, and it would be interesting to explore how the scattering amplitudes \eqref{nto1int}, \eqref{finminus} satisfy them.

\subsection{The limit $p\to 2$}

A time-dependent background similar to the one considered in this work was recently discussed in \cite{Rodriguez:2023kkl,Rodriguez:2023wun} (see also \cite{Collier:2023cyw}). The background in these works is similar to the limit $p\to2$ of our \eqref{defL}, with $\lambda_+=0$, $\lambda_-\not=0$. We focused (in section \ref{timed}) on the opposite case, $\lambda_-=0$, $\lambda_+\not=0$; the two are related by time-reversal symmetry. Interestingly, the results for the correlators in \cite{Rodriguez:2023kkl,Rodriguez:2023wun,Collier:2023cyw} were different from ours. As an example, we found that the $n$-point functions \eqref{absor} vanish, while their analogs in the above papers did not. 

An important difference between the two constructions is that in \cite{Rodriguez:2023kkl,Rodriguez:2023wun}, in our notation, $\lambda_+=0$ and $\lambda_-=-\mu<0$,  see \cite{Collier:2023cyw}. After applying time-reversal, this is related to our system with $\lambda_+=-\mu$, $\lambda_-=0$. We, on the other hand, took $\lambda_+$ to be positive. It is possible that this difference is responsible for the above difference in the correlation functions. Note that flipping the sign of $\lambda_+$ in our construction has an important effect. For positive $\lambda_+$ (our analysis), the potential $V_{\rm st}$, \eqref{sppot}, goes to infinity everywhere in the shaded region in figure \ref{2powa}. For negative $\lambda_+$, there is a region along the positive $v$ axis, where the potential remains small. Thus, incoming $\mathcal{T}^+$ waves can penetrate the potential in this direction, and it is possible that these are the processes captured by the correlation functions of \cite{Rodriguez:2023kkl,Rodriguez:2023wun,Collier:2023cyw}. 

Note also that in the limit $p\to2$, the worldsheet theory \eqref{SwsL}, \eqref{defL}, with $\lambda_+\lambda_-=0$ factorizes into a direct product of a theory for $\phi$ and one for $t$. The $\phi$ theory is a Liouville theory with $c=25$, while that of $t$ is timelike Liouville with $c=1$. This factorization played an important role in the analysis of \cite{Rodriguez:2023kkl,Rodriguez:2023wun,Collier:2023cyw}. For $p<2$, the subject of our paper, this factorization breaks down, and one cannot use the techniques of \cite{Rodriguez:2023kkl,Rodriguez:2023wun,Collier:2023cyw} to analyze the resulting background. This is one of the reasons we used matrix model results to analyze the dynamics. 

We also note that if $\lambda_\pm$ are both non-zero, it looks superficially from \eqref{SwsL}, \eqref{defL} that for $p=2$ the worldsheet theory still factorizes into a product of theories for $\phi$ and $t$, but in fact this is not the case. The reason is that the would be theory for $t$ is a Wick rotated version of the Sine-Gordon model, and the coupling $\lambda_+\lambda_-$ is in this case marginal but not truly marginal. Thus, the $t$ theory exhibits an RG flow, and after coupling to $\phi$ this RG happens as a function of $\phi$,  \cite{Hsu:1992cm}.

\subsection{Closed string radiation}

The time-dependent background considered in this work is reminiscent of the rolling tachyon solution \cite{Sen:2004nf}, which is an open-string analogue of our time-dependent background for $p=2$. However, while in our case the fate of the background is not known, in the rolling tachyon case the brane decays by emitting closed strings. In particular, it was shown in \cite{Lambert:2003zr} that the closed string radiation produced by the rolling tachyon is given by a coherent state of the form
\ie
e^{\alpha a^\dagger}\ket{0}~,
\label{coh_state}
\fe
where $a^\dagger$ is the creation operator for a closed string, and $\alpha$ is proportional to the disk 1-point function for the closed string vertex operator, with rolling tachyon boundary condition. The closed string radiation state produced by the time-dependent background in our case is similar, except now $\alpha$ is computed from the sphere 1-point function \eqref{1ptLo} (for $p<1$), and in particular is of order $1/g_s$. 

From \eqref{coh_state}, we can follow \cite{Lambert:2003zr} and compute the total number $N$ and energy $E$ of closed strings produced by the time-dependent background. We find
\ie
N=\int_0^\infty d\omega ~\omega \left|\big\langle \mathcal{T}^-_{\omega'}\big\rangle_{\lambda_+}\right|^2~,
\\
E=\int_0^\infty d\omega ~\omega^2 \left|\big\langle \mathcal{T}^-_{\omega'}\big\rangle_{\lambda_+}\right|^2~,
\label{NE}
\fe
where the polynomial factors of $\omega$ relative to \cite{Lambert:2003zr} come from the normalization of the tachyon operator, \eqref{voL}. Using the expression \eqref{1ptLo} in \eqref{NE}, we find that the integral converges\footnote{The expressions \eqref{NE} have IR divergences from the small $\omega$ region. These can be dealt with by introducing an IR cutoff.} at large $\omega$ for $p<1$. Note however that $N,~E$ are of order $1/g_s^2$, since $\big\langle \mathcal{T}^-_{\omega'}\big\rangle_{\lambda_+}\sim 1/g_s$. This means that the total number and energy of closed strings produced by the time-dependent background is of the same order as those of the original background, so backreaction should be taken into account. We leave a more detailed understanding of this backreaction to future work.

\subsection{Moving mirror}

Our description of the dynamics in terms of scattering of tachyons off a time-dependent Liouville wall is reminiscent of the moving mirror problem in QFT ~\cite{birrell_davies_1982}. That model is often used as a toy model of black hole physics. In that application, the mirror is taken to be receding from the observer, and approaching the speed of light at late times \cite{Davies:1977yv,Carlitz:1986nh,Almheiri:2013wka,Akal:2020twv}. More generally, the mirror can be taken to move away or towards the observer, and have a final speed different from that of light; see \EG \cite{Akal:2022qei} for a recent discussion. 

In our model, the Liouville wall is qualitatively similar to a moving mirror. The analogy is not precise, since in the case of the moving mirror one typically takes the quantum fields to vanish at the mirror, whereas the Liouville wall is soft, as mentioned earlier in the paper. Another difference between the two problems is that since the Liouville wall is not a physical object, it can move faster than light, something that is not possible for a physical mirror. 

Nevertheless, it would be interesting to study the relation between our results and the moving mirror problem, for example by comparing the particle production and scattering amplitudes we get to those obtained by studying a moving mirror that follows a similar trajectory. This too will be left for future work.

\section*{Acknowledgements}

We thank Ahmed Almheiri, Amit Giveon, Finn Larsen, Massimo Porrati, Victor A. Rodriguez, Zixia Wei, and  Xi Yin for discussions. This work was supported in part by DOE grant DE-SC0009924 and BSF grant 2018068. The work of BB was also supported by  NSF grants PHY-2210349, PHY-2210420, and PHY-2112839.

%%%%%%%%%%%%%%%%%%%%%%%%%%%%%%%%%%%%%%
%%%%%%%%%%%%%%%%%%%%%%%%%%%%%%%%%%%%%%
\appendix
\section{Matrix model dual of $1+1$ dimensional string theory}\label{ap0}
To calculate the correlation functions \eqref{corf}, we use the matrix quantum mechanics dual to $1+1$ dimensional string theory, following the approach of
\cite{Moore:1991zv,Moore:1992ga}. In this appendix, we briefly review this approach.

The Hamiltonian of the matrix quantum mechanics is 
\ie
\label{freeH}
H=\frac{1}{2}\mathrm{Tr}\left(P^2-X^2\right)\ ,
\fe
where $X$ is an $N$-by-$N$ Hermitian matrix and $P$ its canonically conjugate momentum. Closed string excitations are dual to states in the singlet sector of the matrix quantum mechanics. To study that sector, it is convenient to diagonalize $X$ by writing $X=U^{-1} \Lambda U$, where $U\in U(N)$, and $\Lambda=\mathrm{diag}(\lambda_1,\cdots,\lambda_N)$. The singlet sector wavefunction $\Psi$ depends only on the eigenvalues $\lambda_i$, and is completely symmetric under exchange of any pair of eigenvalues, 
\ie
\Psi(\cdots,\lambda_i,\cdots,\lambda_j\cdots)=\Psi(\cdots,\lambda_j,\cdots,\lambda_i\cdots)\ .
\fe
It is useful to make a similarity transformation, $H=\Delta \widetilde H \Delta^{-1}$, where $\Delta=\prod_{i<j}^N(\lambda_i-\lambda_j)$ is the Vandermonde determinant, and 
\ie
\widetilde H=\frac{1}{2}\sum_{i=1}^N\left(-\partial_{\lambda_i}^2-\lambda_i^2\right)\ .
\fe
The Hamiltonian $\widetilde H$ acts on the wavefunction $\widetilde \Psi(\lambda_i)=\Delta \Psi(\lambda_i)$. The wavefunction $\widetilde \Psi(\lambda_i)$ is antisymmertic under exchanging any pair of eigenvalues. Thus, the Hamiltonian $\widetilde H$ describes a system of $N$ non-relativistic, non-interacting fermions, moving in the potential $V(\lambda)=-\frac{\lambda^2}{2}$.

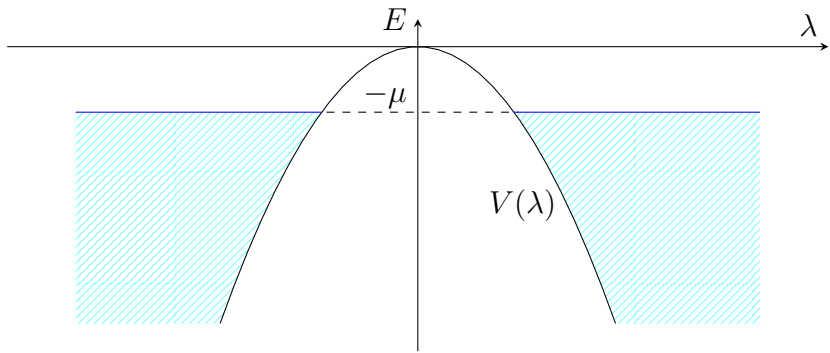
\begin{figure}[t]
\centering
{
\begin{tikzpicture}
\begin{axis}[axis lines=middle,
			 ylabel style={at=(current axis.above origin), anchor=east},
            xlabel=$\lambda$,
            ylabel=$E$,
            enlargelimits,
            ytick=\empty,
            xtick=\empty,
            height=6cm,
            width=12.5cm
           ]
%\node[coordinate,pin=0:{$-\mu$}] at (axis cs:0,-1){};

\node[left,black] at (axis cs:0,-0.15){$-\mu$};
\addplot[mark=none, dashed,domain={-0.6:0.662},black,samples=3] {-0.2};

\addplot[black,domain={2.236:2.2359},samples=2] {-0.5*x^2+0.1*x^4} node[pos=0.85, left]{};
\addplot[name path=F,black,domain={-1.3:1.3},samples=30] {-0.5*x^2} node[pos=0.85, left]{$V(\lambda)$};
\addplot[name path=G,mark=none,domain={0.632456:2.25},blue,samples=2] {-0.2};
\addplot[name path=H,mark=none,domain={-2.25:-0.632456},blue,samples=2] {-0.2};

\path[name path=neg] (axis cs:0,-0.845) -- (axis cs:3,-0.845);
\path[name path=neg2] (axis cs:-3,-0.845) -- (axis cs:0,-0.845);
\addplot[pattern=north east lines, pattern color=cyan!50]fill between[of=F and G, soft clip={domain=0.632456:1.3}];
\addplot[pattern=north east lines, pattern color=cyan!50]fill between[of=neg and G, soft clip={domain=1.3:2.25}];
\addplot[pattern=north east lines, pattern color=cyan!50]fill between[of=F and H, soft clip={domain=-1.3:-0.632456}];
\addplot[pattern=north east lines, pattern color=cyan!50]fill between[of=neg2 and H, soft clip={domain=-2.25:-1.3}];
\end{axis}
\end{tikzpicture}
}
\caption{Semiclassical description of the closed string vacuum in the dual $c=1$ matrix quantum mechanics.}
\label{vacuum}
\end{figure}

\begin{figure}[t]
\centering
 \includegraphics[width=0.45\textwidth]{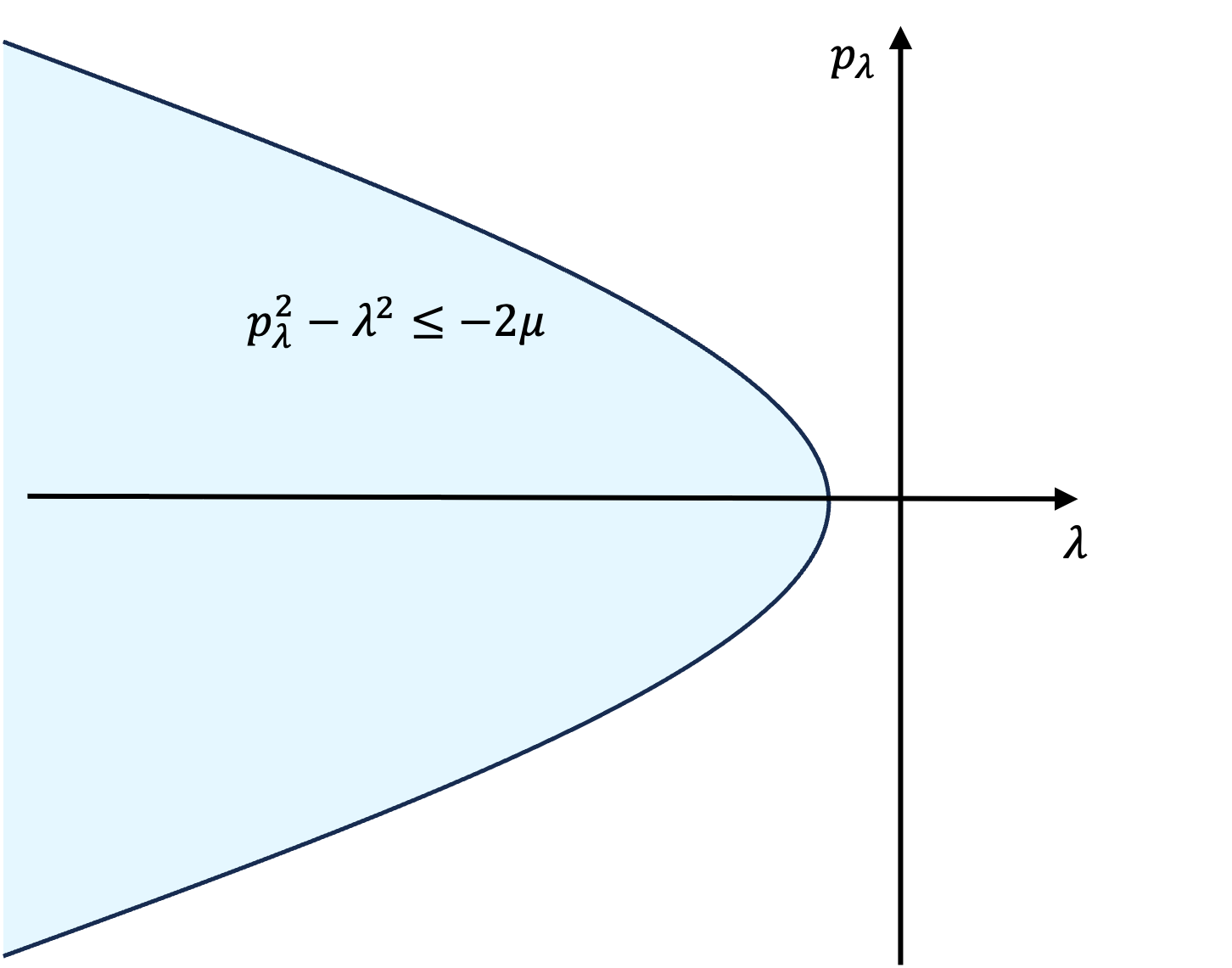}
 \caption{The Fermi sea (shaded region) corresponding to the standard background  of $1+1$ dimensional string theory described in section \ref{Euc}.} 
\label{fes}
\end{figure}
The potential $V(\lambda)$ is unbounded from below. To arrive at the theory that is dual to two-dimensional string theory, we consider the following {\textit{double-scaling limit}}. We study the system at fixed Fermi energy $E=-\mu<0$, and send $N\to\infty$. Thus, the fermions fill all states with energy less than $E=-\mu$; see figure \ref{vacuum}. This is the state that is dual to the closed string vacuum. 

The vertex operators \eqref{vo} correspond to infinitesimal perturbations of the closed string vacuum. It is sufficient to consider perturbations of the Fermi surface on one side of the potential in the figure, say the one with $\lambda<0$. As is clear from figure \ref{vacuum}, effects that mix the two sides are non-perturbative in $1/\mu$. They correspond to non-perturbative effects in string theory; see \EG 
\cite{Balthazar:2019rnh,Balthazar:2019ypi,Sen:2019qqg,Sen:2020eck,Sen:2021qdk,Balthazar:2022apu,Alexandrov:2023fvb} for recent discussions.

In a semiclassical description, the fermions fill a Fermi sea up to the surface $\lambda=\sqrt{p_{\lambda}^2+2\mu}$, where $p_{\lambda}$ denotes the momentum conjugate to $\lambda$ (see figure \ref{fes}). The string coupling $g_s$ is given by $2\pi g_s\equiv\mu^{-1}$, while closed string excitations are dual to perturbations of the Fermi surface; see\cite{Ginsparg:1993is,Jevicki:1993qn,Polchinski:1991uq} for details.

The Fermi surface can be written in terms of the fermion density, which is a fermion bilinear. Using this, correlation functions of tachyon operators \eqref{vo} can be calculated using the dual matrix quantum mechanics description. This formalism allows one to express the string theory correlation functions \eqref{corf} in terms of the S-matrix of free fermions \cite{Moore:1991zv}.

By decomposing the fluctuations of the Fermi sea into particle-hole pairs, \cite{Moore:1991zv} developed an efficient formalism for computing the S-matrix $\mathcal{R}(q_j,-q_l')$ ($q_j,q'_l>0$). For a particle/hole pair with Euclidean momentum $q$, the reflection amplitude is
\ie
\label{ream}
R_q=\mu^{-|q|}\sqrt{\frac{2}{\pi}}e^{i\pi/4}\cos \left(\frac{\pi}{2}(\frac{1}{2}+i\mu-|q|)\right)\Gamma\left(\frac{1}{2}-i\mu+|q|\right)\ .
\fe
In terms of the reflection amplitude, the S-matrix is constructed as follows.
\ie
\label{Rqq}
\mathcal{R}(q_j,-q_l')=&\delta \left(\sum_j^nq_j-\sum_l^{n'}q'_l\right)i^{n+n'}\sum_{k=1}^{\min\{n,n'\}}\frac{1}{k}\sum_{\text{AF}_k}\int dQ \\
&\prod_{j=2}^{k+1}\bigg[\sum_{T_+\subseteq F_{2j}-F_{2j-1}}(-1)^{|T_+|}\theta\Big(q(T_+)-\big(Q+q(F_{2j})\big)\Big)\bigg]\\
&\prod_{j=1}^k\bigg[\sum_{T_-\subseteq F_{2j+1}-F_{2j}}(-1)^{|T_-|}\theta\Big(Q+q(F_{2j})+q(T_-)\Big)\bigg]\\
&\prod_{j=1}^kR_{Q+q(F_{2j+1})}R^*_{Q+q(F_{2j+2})}\ ,
\fe
where $q_j$ denote the incoming Euclidean momenta, $-q'_\ell$ denote the outgoing Euclidean momenta, $q(F_j)$ denotes the sum of the momenta in the set $F_j$, and AF$_k$ is an admissible filtration (AF) of order $k$ defined as a tower of subsets, \IE
\ie
\varnothing \equiv F_2\subset F_3\subset\cdots\subset F_{2(k+1)}\equiv \{q_j,-q'_l\}\ ,
\fe
such that $F_{2j+1}-F_{2j}$ is a set of negative momenta, namely
\ie
F_{2j+1}-F_{2j}\subseteq \{-q_1',-q_2',\cdots,-q_{n'}'\}~,
\fe
and $F_{2j}-F_{2j-1}$ is a set of positive momenta, namely
\ie
F_{2j}-F_{2j-1}\subseteq \{q_1,q_2,\cdots,q_n\}~.
\fe
The theta functions in \eqref{Rqq} restricts the integration range of the integration variable $Q$, namely for all $j$ we have
\ie
\label{ineQ}
-q(T_-)<Q+q(F_{2j})<q(T_+)\ .
\fe
Since $T_+\subseteq F_{2j}-F_{2j-1}$,
\ie
Q+q(F_{2j})<q(F_{2j})-q(F_{2j-1})\ ,
\fe
which is equivalent to $Q+q(F_{2j-1})<0$. So, from the definition \eqref{ream},
\ie
R_{Q+q(F_{2j-1})}=&\mu^{Q+q(F_{2j-1})}\sqrt{\frac{2}{\pi}}e^{i\pi/4}\cos \left(\frac{\pi}{2}(\frac{1}{2}+i\mu+Q+q(F_{2j-1}))\right)\\
&\Gamma\left(\frac{1}{2}-i\mu-Q-q(F_{2j-1})\right)\ .
\fe
On the other hand, in \eqref{ineQ} $T_-\subseteq F_{2j+1}-F_{2j}$, which means that $q(T_-)<0$. Therefore, $Q+q(F_{2j})>0$ and
\ie
R^*_{Q+q(F_{2j})}=&\mu^{-Q-q(F_{2j})}\sqrt{\frac{2}{\pi}}e^{-i\pi/4}\cos \left(\frac{\pi}{2}(\frac{1}{2}-i\mu-Q-q(F_{2j}))\right)\\
&\Gamma\left(\frac{1}{2}+i\mu+Q+q(F_{2j})\right)\ .
\fe
Combining the two reflection amplitudes, we find the interesting property that 
\ie
\label{convert}
\mu^{\sum_{j=1}^{n}q_j}\prod_{j=1}^kR_{Q+q(F_{2j+1})}R^*_{Q+q(F_{2j+2})}
\fe
depends on $\mu$ and $Q$ only through the combination $i\mu+Q$. 

The matrix model S-matrix $\mathcal{R}(q_j,-q_l')$ is dual to string theory scattering amplitude of closed strings. The relation between them is given by
\ie
\label{corrR}
\Big\langle \prod_{j=1}^nT_{q_j}\prod_{l=1}^{n'}T_{-q'_l}\Big\rangle=\frac{\mu^{\sum_{j=1}^nq_j}\mathcal{R}(q_j,-q_l')}{\prod_{j=1}^nq_j\prod_{l=1}^{n'}q'_l}~.
\fe
In practice, it is easier to calculate scattering amplitudes with an insertion of the cosmological constant operator $T_0$. As discussed below \eqref{corf}, the insertion of $T_0$ is equivalent to acting with $-\partial_\mu$ on the correlation function with $T_0$ stripped off. Thus, we have
\ie
\label{epscorr}
    \Big\langle T_0\prod_{j=1}^nT_{q_j}\prod_{l=1}^{n'}T_{-q'_l}\Big\rangle=&\lim_{\epsilon\to 0^+}\Big\langle T_{n\epsilon}\prod_{j=1}^nT_{q_j-\epsilon}\prod_{l=1}^{n'}T_{-q'_l}\Big\rangle\\
    =&-\lim_{\epsilon\to 0^+}\partial_\mu\Big\langle \prod_{j=1}^nT_{q_j-\epsilon}\prod_{l=1}^{n'}T_{-q'_l}\Big\rangle\\
    =&-\lim_{\epsilon\to 0^+}\frac{1}{ \prod_{j=1}^nq_j\prod_{l=1}^{n'}q'_l }\frac{\partial}{\partial \mu}\left(\mu^{\sum_{j=1}^nq_j}\mathcal{R}(q_j-\epsilon,-q_l')\right)\ .
\fe
Then, since from \eqref{Rqq} and \eqref{convert} $\mathcal{R}(q_j,-q_l')$ depends on $\mu$ and $Q$ only through the combination $i\mu+Q$, it follows that we can convert the $\mu$-derivative into a $Q$-derivative. More explicitly,
\ie
\label{sumaf}
\begin{split}
\frac{\partial}{\partial \mu}&\left(\mu^{\sum_{j=1}^nq_j}\mathcal{R}(q_j-\epsilon,-q_l')\right)=\delta \left(\sum_j^n(q_j-\epsilon)-\sum_l^{n'}q'_l\right)i^{n+n'+1}\mu^{\sum_{j=1}^nq_j} \\
&\sum_{k=1}^{\min\{n,n'\}}\frac{1}{k}\sum_{\text{AF}_k}\int dQ\Bigg\{\partial_Q\bigg[\prod_{j=2}^{k+1}\sum_{T_+\subseteq F_{2j}-F_{2j-1}}(-1)^{|T_+|}\theta\Big(q(T_+)-\big(Q+q(F_{2j})\big)\Big)\\
&\prod_{j=1}^k\sum_{T_-\subseteq F_{2j+1}-F_{2j}}(-1)^{|T_-|}\theta\Big(Q+q(F_{2j})+q(T_-)\Big)\bigg]\prod_{j=1}^kR_{Q+q(F_{2j+1})}R^*_{Q+q(F_{2j+2})}\Bigg\}\ ,
\end{split}
\fe
where we have integrated by parts in $Q$. The derivative of the theta functions are just delta functions, which after the integration give a finite sum. 

Finally, note that in the compact case $X\sim X+2\pi R$, the Dirac delta function for momentum conservation should be replaced by a Kronecker delta times $2\pi R$.

\section{Closed string amplitudes}\label{pp}

In this appendix, we outline the calculations of the closed string amplitudes, $\langle T_0T_p^m\prod_{j=1}^nT_{-q_j}\rangle$ using the matrix model techniques discussed in appendix \ref{ap0}.

\subsection{One outgoing particle}\label{pp1}
For $q=mp$ ($m\in Z^+$), \eqref{epscorr} and \eqref{sumaf} lead to
\ie
\label{corrq}
\langle T_0T_p^mT_{-q}\rangle=&\lim_{\epsilon\to 0^+}\langle T_\epsilon T_{p-\epsilon}^mT_{-q}\rangle
\\
=&\lim_{\epsilon\to 0^+}2\pi R\frac{i^m}{p^mq }\mu^{mp}\int dQ \Bigg\{\partial_Q\bigg[\sum_{T_+\subseteq F_4-F_3}(-1)^{|T_+|}\theta\Big(q(T_+)-\big(Q+q(F_{4})\big)\Big)\\
&\sum_{T_-\subseteq F_3-F_2}(-1)^{|T_-|}\theta\Big(Q+q(F_4)+q(T_-)\Big)\bigg]R_{Q+q(F_3)}R^*_{Q+q(F_4)}\Bigg\}\ ,
\fe
where
\ie
F_3-F_2=\{-q\}\ ,
\fe
and
\ie
F_4-F_3=\{\underbrace{p-\epsilon,\cdots,p-\epsilon}_m\}\ .
\fe

Performing the integration over $Q$ we find,
\ie
&\langle T_0T_p^mT_{-q}\rangle= \lim_{\epsilon\to 0^+}2\pi R\frac{i^m}{p^mq }\mu^{mp} \\
&\Bigg[-\sum_{T_+\subseteq F_4-F_3}(-1)^{|T_+|}R_{q(T_+)-mp}R^*_{q(T_+)}\sum_{T_-\subseteq F_3-F_2}(-1)^{|T_-|}\theta\Big(q(T_+)+q(T_-)\Big)\\
&+\sum_{T_-\subseteq F_3-F_2}(-1)^{|T_-|}R_{-q(T_-)-mp}R^*_{-q(T_-)}\sum_{T_+\subseteq F_4-F_3}(-1)^{|T_+|}\theta\Big(q(T_+)+q(T_-)\Big)\Bigg]\ .
\fe
The first term in the square bracket can be simplified to
\ie
-\sum_{b=0}^m(-1)^b\binom{m}{b}R_{(b-m)p}R^*_{bp}\ .
\fe
The second term in the square bracket vanishes. To see this, consider the two possible choices for $T_-$. If $T_-=\varnothing$,
\ie
\sum_{T_+\subseteq F_4-F_3}(-1)^{|T_+|}\theta\Big(q(T_+)+q(T_-)\Big)=&\sum_{T_+\subseteq F_4-F_3}(-1)^{|T_+|}\theta\Big(q(T_+)\Big)
=\sum_{b=0}^m(-1)^b\binom{n}{b}
=0\ .
\fe
On the other hand, if $T_-=\{-q\}$, $q(T_+)+q(T_-)=q(T_+)-q\le -m\epsilon<0$. So, the theta function gives zero. Therefore,
\ie
\label{TqTpn}
\langle T_0T_p^mT_{-q}\rangle=-2\pi R \frac{i^m}{p^mq }\mu^{mp}\sum_{b=0}^m(-1)^b\binom{m}{b}R_{(b-m)p}R^*_{bp}\ .
\fe
To leading order in the $1/\mu$ expansion (\IE to leading order in string perturbation theory), we find  
\ie
\label{corrmpq}
\langle T_0T_{p}^mT_{-q}\rangle=&(-1)^{m+1}2\pi R\mu^{mp-m}\prod_{i=1}^{m-1}(q-i)
=2\pi R\mu^{mp-m}\frac{\Gamma(m(1-p))}{\Gamma(1-mp)}~.
\fe
We have checked that \eqref{corrmpq} follows from \eqref{TqTpn} for $1\leq m\leq 20$, and conjecture that the result holds for all integers $m$. In this case, this conjecture is known to be correct, since the scattering amplitude in \eqref{TqTpn} is known exactly, see \EG \cite{Kutasov:1991pv,Ginsparg:1993is}. However, for us this is a warmup exercise towards other cases, where the answer is not known from other work. 

Using \eqref{TqTpn} one can compute corrections to \eqref{corrmpq} in the $1/\mu$ expansion. The leading correction is given by 
\ie
\label{TqTpmtorus}
\langle T_0T_{p}^mT_{-q}\rangle_{\text{torus}}=&-2\pi R\mu^{mp-m}\frac{\Gamma(m(1-p)+2)}{\Gamma(1-mp)}\frac{mp^2-mp-1}{24\mu^2}\ .
\fe

\subsection{Two outgoing particles}\label{pp2}
In this subsection, we calculate the correlation function $\langle T_0T_p^mT_{-q_1}T_{-q_2}\rangle$, for $q_1=m_1p>0$ and $q_2=(m-m_1)p>0$ with integers $0<m_1<m$. First, from \eqref{epscorr} and \eqref{sumaf},
\ie
\label{corrq1q2}
&\langle T_0T_p^mT_{-q_1}T_{-q_2}\rangle\\
=& \lim_{\epsilon\to 0^+}2\pi Ri^{m+1}\mu^{mp}\frac{1}{ p^mq_1q_2 }\sum_{k=1}^2\frac{1}{k}\sum_{\text{AF}_k}\int dQ \prod_{j=1}^k\Bigg\{R_{Q+q(F_{2j+1})}R^*_{Q+q(F_{2j+2})}\\
&\partial_Q\bigg[\prod_{j=2}^{k+1}\sum_{T_+\subseteq F_{2j}-F_{2j-1}}(-1)^{|T_+|}\theta\Big(q(T_+)-\big(Q+q(F_{2j})\big)\Big)\\
&\prod_{j=1}^k\sum_{T_-\subseteq F_{2j+1}-F_{2j}}(-1)^{|T_-|}\theta\Big(Q+q(F_{2j})+q(T_-)\Big)\bigg]\Bigg\}\ .
\fe
When $k=1$,
\ie
F_3-F_2=\{-q_1,-q_2\}\ ,
\fe
and
\ie
F_4-F_3=\{\underbrace{p-\epsilon,\cdots,p-\epsilon}_m\}\ .
\fe
Without loss of generality, we take $m_1<m-m_1$. Then, the contribution of AF$_1$ to the r.h.s. of \eqref{corrq1q2} is
\ie
&\int dQ\Bigg\{-\sum_{b=0}^m(-1)^b\binom{m}{b}\delta\Big(bp-Q\Big)\sum_{T_-\subseteq \{-q_1,-q_2\}}(-1)^{|T_-|}\theta\Big(Q+q(T_-)\Big)R_{Q-mp}R^*_Q\\
&+\sum_{b=0}^m(-1)^b\binom{m}{b}\theta\Big(bp-Q\Big)\sum_{T_-\subseteq \{-q_1,-q_2\}}(-1)^{|T_-|}\delta\Big(Q+q(T_-)\Big)R_{Q-mp}R^*_Q\Bigg\}\\
=&-\sum_{b=0}^{m_1}(-1)^b\binom{m}{b}R_{(m-b)p}R_{bp}^*+\sum_{b=m-m_1+1}^m(-1)^b\binom{m}{b}R_{(m-b)p}R_{bp}^*\\
&-(-1)^{m_1+1}\binom{m-1}{m_1}R_{q_2}R_{q_1}^*-(-1)^{m-m_1+1}\binom{m-1}{m-m_1}R_{q_1}R_{q_2}^*\ .
\fe
When $k=2$ on the r.h.s. of \eqref{corrq1q2}, $F_3$ can be either $\{-q_1\}$ or $\{-q_2\}$. Due to the symmetry of exchanging $F_3-F_2\leftrightarrow F_5-F_4$ and $F_4-F_3\leftrightarrow F_6-F_5$ in \eqref{corrq1q2}, both the possibilities lead to the same contribution. So we only need to look at one of them, namely
\ie
F_{2j+1}-F_{2j}=\{-q_j\}\ .
\fe
Besides,
\ie
F_{2j}-F_{2j-1}=\{\underbrace{p-\epsilon,\cdots,p-\epsilon}_{n_j}\}\ ,
\fe
with any integers $n_1,n_2>0$ such that $n_1+n_2=m$. Then, the contribution of AF$_2$ to the r.h.s. of \eqref{corrq1q2} is given by
\ie
&2\sum_{n_1=1}^{m-1}\frac{m!}{n_1!(m-n_1)!}\Bigg[-(-1)^{n_1+m_1+1}\binom{m-n_1-1}{m_1}R_0R_{q_2-n_1p}R_{q_1}^*R_{n_1p}^*\\
&-\sum_{b=0}^{\min\{m-m_1,n_1-1\}}(-1)^{m_1-n_1+1}\binom{n_1}{b}\binom{m-n_1-1}{m_1-n_1+b}R_{q_2-bp}R_{(n_1-b)p}R_{bp}^*R_{q_1-(n_1-b)p}^*\\
&-\sum_{b=0}^{m_1}(-1)^{n_1-m_1}\binom{m-n_1}{b}\binom{n_1-1}{b+n_1-m_1-1}R_{q_1-bp}R_{(m-n_1-b)p}R_{bp}^*R_{(b+n_1)p-q_1}^*\Bigg]\ .
\fe
Note that the overall factor of 2 comes from the same contribution for the case $F_3=\{-q_2\}$, and it will cancel with the factor $\frac{1}{k}$ in \eqref{corrq1q2}.

Extrapolating from the cases $1\le m\le 12$, we conjecture the following general formula at leading order in $1/\mu$:
\ie
\label{Tq1q2T0}
\langle T_0T_p^mT_{-q_1}T_{-q_2}\rangle=&(-1)^m 2\pi R\mu^{mp-m-1}\frac{m!}{m_1!(m-m_1)!}\prod_{i_1=1}^{m_1}(q_1-i_1)\prod_{i_2=1}^{m-m_1}(q_2-i_2)\\
=& 2\pi R\mu^{mp-m-1}m!\frac{\Gamma\left((1-p)m_1+1\right)}{\Gamma\left(m_1+1\right)\Gamma(1-pm_1)}\frac{\Gamma\left((1-p)m_2+1\right)}{\Gamma\left(m_2+1\right)\Gamma(1-pm_2)}\ .
\fe
The leading  correction to \ref{Tq1q2T0} is
\ie
\label{Tq1q2Tpmtorus}
\langle T_0T_p^mT_{-q_1}T_{-q_2}\rangle_{\text{torus}}=&\langle T_0T_p^mT_{-q_1}T_{-q_2}\rangle_{\text{sphere}}\frac{mp-m-2}{24\mu^2}\left(m+1+(m^2-m_1m_2)p\right.\\
&\left.+(-2m^2-m+2m_1m_2)p^2+(m^2-m_1m_2)p^3\right)\ .
\fe

\subsection{Three and four outgoing particles}

Let's now consider the amplitude for three outgoing particles, $\langle T_0T_p^m\prod_{j=1}^3T_{-q_j}\rangle$. Using \eqref{epscorr} and \eqref{sumaf}, let $m_j=q_j/p$ be integers and $m=m_1+m_2+m_3$. To leading order in  $1/\mu$ we find
\ie
\label{3pt0p}
\langle T_0T_p^mT_{-q_1}T_{-q_2}T_{-q_3}\rangle=&-(-1)^m2\pi R\mu^{mp-m-2}m!(mp-m-1)\prod_{j=1}^3\frac{\prod_{i=1}^{m_j}(q_j-i)}{m_j!}\\
=&-2\pi R\mu^{mp-m-2}m!(mp-m-1)\prod_{j=1}^3\frac{\Gamma\left((1-p)m_j+1\right)}{\Gamma\left(m_j+1\right)\Gamma(1-pm_j)}\ ,
\fe
where we checked this result for $1\le m\le 10$ and conjectured the general formula.

For four outgoing particles, $\langle T_0T_p^m\prod_{j=1}^4T_{-q_j}\rangle$, we again start  with $q_j=m_jp$, $m_j\in Z_+$, where $m_1+m_2+m_3+m_4=m$. We find
\ie
&\langle T_0T^m_pT_{-q_1}T_{-q_2}T_{-q_3}T_{-q_4}\rangle\\
=& (-1)^m2\pi R\mu^{mp-m-3}m!(mp-m-1)(mp-m-2)\prod_{j=1}^4\frac{\prod_{i=1}^{m_j}(m_jp-i)}{m_j!}\\
=& 2\pi R\mu^{mp-m-3}m!(mp-m-1)(mp-m-2)\prod_{j=1}^4\frac{\Gamma\left((1-p)m_j+1\right)}{\Gamma\left(m_j+1\right)\Gamma(1-pm_j)}\ ,
\fe
where we checked the result for $1\le m=m_1+m_2+ m_3+m_4\le 10$.

\subsection{Any number of outgoing particles}

Interestingly, from the results of $\langle T_0T_p^m\prod_{j=1}^nT_{-q_j}\rangle$ with $n=1,2,3,4$, we find that the correlation functions can be expressed in a universal way:
\ie
\langle T_0T^m_p\prod_{j=1}^nT_{-q_j}\rangle=(-1)^n2\pi R\partial_\mu^{n-2}\mu^{mp-m-1}m!\prod_{j=1}^n\frac{\Gamma\left((1-p)m_j+1\right)}{\Gamma\left(m_j+1\right)\Gamma(1-pm_j)}~.
\fe
Integrating w.r.t. $\mu$ once, we can get rid of $T_0$ by using $T_0=-\partial_\mu$, to find
\ie
\label{npt}
\langle T^m_p\prod_{j=1}^nT_{-q_j}\rangle=(-1)^{n-1}2\pi R\partial_\mu^{n-3}\mu^{mp-m-1}m!\prod_{j=1}^n\frac{\Gamma\left((1-p)m_j+1\right)}{\Gamma\left(m_j+1\right)\Gamma(1-pm_j)}\ .
\fe
We conjecture that this result holds for all $n$.

\bibliographystyle{JHEP}
\bibliography{sinegordon}
%%%%%%%%%%%%%%%%%%%%%%%%%%%%%%%%%%%%%%
%%%%%%%%%%%%%%%%%%%%%%%%%%%%%%%%%%%%%%

\end{document}